\documentclass[useAMS,usenatbib,usegraphicx]{mn2e}

\newcommand{\Revised}[1]{#1}

\newcommand{\be}{\begin{equation}}
\newcommand{\ee}{\end{equation}}
\newcommand{\bea}{\begin{eqnarray}}
\newcommand{\eea}{\end{eqnarray}}

\newcommand{\dd}{{\rm d}}

\newcommand{\mm}{_{\rm max}}
\newcommand{\Vm}{V\mm}
\newcommand{\Rm}{R\mm}

\newcommand{\Vc}{V_{\rm c}}
\newcommand{\rr}{_{\rm r}}
\newcommand{\vir}{_{\rm vir}}

\newcommand{\hMsol}{{\,\rm h^{-1}\,M}_\odot}
\newcommand{\kpc}{{\,\rm kpc}}
\newcommand{\Mpc}{{\,\rm Mpc}}
\newcommand{\hMpc}{\,h^{-1}\Mpc}
\newcommand{\hkpc}{\,h^{-1}\kpc}

\title{The dynamical structure of dark matter haloes}

\author[Y. Ascasibar and S. Gottl\"{o}ber]
{
  Yago Ascasibar\thanks{E-mail: yago@aip.de} and Stefan Gottl\"{o}ber\\
  Astrophysikalisches Institut Potsdam, An der Sternwarte 16, 14482 Potsdam, Germany
}

\date{Draft version -- \today}

\begin{document}
\maketitle

\begin{abstract}

Thanks to the ever increasing computational power and the development of more sophisticated algorithms, numerical $N$-body simulations are now uncovering several phenomenological relations between the physical properties of dark matter haloes in position and velocity space.
It is the aim of the present work to investigate in detail the dynamical structure of dark matter haloes, as well as its possible dependence on mass and its evolution with redshift up to $z=5$.
We use high-resolution cosmological simulations of individual objects to compute the radially-averaged profiles of several quantities, scaled by the radius $\Rm$ at which the circular velocity attains its maximum value, $\Vm$.
We investigate the circular velocity profile, the dark matter density and its logarithmic slope, the position and velocity of the centre of mass on different scales, the radial infall around the halo, its spin parameter, the radial and tangential components of the velocity dispersion, and the coarse-grained phase-space density.
It is found that all the physical properties considered display a similar structure when expressed in terms of $\Rm$ and $\Vm$.
No systematic dependence on mass or cosmic epoch are found within $\Rm$, and all the different radial profiles are well fit by simple analytical models.
However, our results suggest that several properties are not `universal' outside this radius.
In particular, dark matter haloes should not be assumed to be in equilibrium beyond $\Rm$, especially at high redshift, where significant infall velocities can be measured.
We therefore conclude that the whole dynamical structure of haloes, rather than just their density profile, is `universal' within $\Rm$ at least up to $z=5$,
\Revised{
and that some dynamical process yet to be identified is responsible for such universality, rather than the details of the merger history of the halo.
Memory of the initial conditions is only retained in the outer regions, where there is significant scatter from object to object, as well as clear systematic trends with mass and time.
Thus, we argue that $\Rm$ and $\Vm$ provide a well-motivated, model-independent choice to characterize dark matter haloes, especially in comparison with the so-called `virial' mass or radius.
}
\end{abstract}


  \section{Introduction}
  \label{secIntro}

The internal structure of dark matter haloes is one of the key predictions of any model of structure formation.
Although the complexity of the non-linear phase of gravitational collapse has so far precluded an exact analytical treatment, much progress has been possible over the last three decades thanks to cosmological numerical experiments \citep[see e.g.][for a historical overview]{Bertschinger98}.

Indeed, perhaps one of the most well-known results of $N$-body simulations is the so-called `universality' of the dark matter density profile, $\rho(r)$, of the simulated haloes \citep{NFW97}.
Although the asymptotic behaviour towards the centre \citep[see e.g.][and references therein, for a recent discussion]{Merritt+06} or the outer parts \citep{Prada+06} are still open questions, there is general consensus in that the radial density profile of dark matter haloes can be well fit by a simple analytical function with very few free parameters.
Many of the functional forms proposed in the literature can be cast in the form
\be
\rho(r)=\frac{\rho_0}{x^\alpha \left[1+(r/r_0)^\beta\right]^\gamma}
\ee
where the characteristic density and radius $\rho_0$ and $r_0$ are usually regarded as free parameters, and different prescriptions exist for the exponents $\alpha$, $\beta$, and $\gamma$.
It is important to note that $\rho_0$ and $r_0$ have been found to be correlated, a phenomenon referred to as the `mass-concentration relation' \citep[e.g.][]{Bullock+01_cM,Eke+01,Zhao+03b,Colin+04,Maccio+07,Neto+_07}, and it has been claimed that, for any
 given object, the product $\rho_0 r_0$ stays approximately constant throughout its evolution \citep{Romano+07}.

In velocity space, dark matter haloes also seem to display a remarkable degree of universality, with the angular momentum \citep[e.g.][]{BarnesEfstathiou87,Warren+92,ColeLacey96,Bullock+01_j,Peirani+04,BailinSteinmetz05,HetzneckerBurkert06,GottloeberYepes07}, velocity dispersion \citep[e.g][]{Rasia+04,DehnenMcLaughlin05,Merritt+06} and anisotropy \citep[e.g.][]{ColeLacey96,Carlberg+97,Colin+00,FukushigeMakino01,Rasia+04,DehnenMcLaughlin05} profiles being well described by simple analytical fits at $z=0$.

Of course, the density profile and the velocity structure of the haloes are not independent, and different correlations have been observed.
For instance, a linear relation between the anisotropy parameter $\beta$ and the local slope of the density profile has recently been advocated \citep{HansenMoore06,HansenStadel06}, and it has been argued \citep{TaylorNavarro01} that the coarse-grained phase-space density profile, $q(r)$, of galaxy-sized haloes follows a power law
\be
q(r) \equiv \frac{\rho(r)}{\sigma^3(r)} \propto r^{-\alpha}
\label{eqTN01}
\ee
 with exponent $\alpha=15/8=1.875$.
This result was confirmed by \citet{Rasia+04} on the scale of galaxy clusters, where the value $\alpha=1.95$ was found to provide the best fit to numerical data.
\citet{Ascasibar+04} showed that equation~(\ref{eqTN01}) actually holds at all scales, from galaxies to galaxy clusters, when expressed in units of `virial' quantities,
\be
q(r) = K \frac{\rho_{\rm c}}{V\vir^3} \left(\frac{r}{R\vir}\right)^{-\alpha}
\ee
where $\rho_{\rm c}$ denotes the critical density of the universe, $M\vir$ and $R\vir$ are the virial mass and radius of the halo, $V\vir^2=GM\vir/R\vir$, $K=10^{1.46\pm0.04}$, and $\alpha=1.90\pm0.05$.
The scaling of $q$ with halo mass or velocity dispersion has also been studied by several authors \citep{DalcantonHogan01,Dave+01,Peirani+06}, and the evolution of this quantity during the merging history of the halo has been addressed by \citet{Peirani+06} and \citet{Hoffman+_07}.
Using the Jeans equation, \citet{TaylorNavarro01} computed the density profile expected for systems in equilibrium that fulfill equation~(\ref{eqTN01}).
Families of analytical solutions have later been discussed in detail for both the isotropic \citep{Hansen04,Austin+05,Barnes+06} and the anisotropic case \citep{DehnenMcLaughlin05,Barnes+07}.

To summarize, there is plenty of evidence suggesting that the present-day dynamical structure of dark matter haloes may be quite insensitive to the details of their accretion history.
More precisely, the radially-averaged profiles of
\begin{enumerate}
\item density (or mass)
\item angular momentum
\item velocity dispersion (radial and tangential)
\item coarse-grained phase-space density
\end{enumerate}
are similar for all haloes, and can be described in terms of a few free parameters.
In this paper, we use a set of high-resolution cosmological N-body simulations to investigate the dependence of the dynamical structure on halo mass and epoch, up to $z=5$.
Apart from the quantities listed above, we also consider the radial infall velocity and the offset between the centres of mass on different scales.
It is shown that, at all redshifts, all the relevant properties of dark matter haloes can be expressed in terms of the characteristic radius $\Rm$ at which the circular velocity, $\Vc^2=GM/r$, attains its maximum value, $\Vm$.
These two parameters are model-independent, they can be easily measured for any dark matter halo, and they provide a reasonable indication of the extent of the inner region where the halo is in equilibrium.

The paper is structured as follows: our numerical experiments are briefly described in Section~\ref{secSims}.
Results are presented in Section~\ref{secResults}, and the main conclusions are summarized in Section~\ref{secConclus}.

  \section{Numerical experiments}
  \label{secSims}

The present work is based on two sets of $N$-body simulations carried out with the Adaptive Refinement Tree code \citep{Kravtsov+97_ART}.
Both of them are random realizations of a $\Lambda$CDM cosmology with parameters $\Omega_{\rm m}=0.3$, $\Omega_{\Lambda}=0.7$, $h=0.7$, and $\sigma_8 = 0.9$, and the box size is $L=80\hMpc$.
In one case, we have selected eight cluster-sized haloes at $z=0$ to be resimulated by the multiple-mass technique \citep{Klypin+01}, while six galaxy-sized haloes have been selected in the other set.
Particle masses are $m_{\rm p}=3.16\times10^8$ and $5.0\times10^6\hMsol$, respectively, and the force resolution reached (two times the size of the highest refinement level cell) was $2.4$ and $0.6\hkpc$ in each case.
Mass and force resolution are an important concern in order to obtain reliable measurements of the internal structure of dark matter haloes.
At high redshift, mass resolution becomes especially critical, since even the most massive objects are composed by a small number of particles.

The mass distribution of this sample of 14 simulated haloes has been investigated in \citet{Ascasibar+07}, where it was compared with the predictions of the spherical infall model.
Here we extend the analysis to the internal structure of these objects in velocity space.
It is interesting to stress that our haloes have been randomly selected, i.e. no effort has been made to select isolated nor dynamically relaxed systems, and therefore they represent an unbiased sample of the average cosmic population.

\begin{table*}
\centerline{
\begin{tabular}{rrrrrrrrr}
\hline\hline
\multicolumn{3}{c}{$z=0$} & \multicolumn{3}{c}{$z=1$} & \multicolumn{3}{c}{$z=5$} \\
 $\Rm~~$ & $\Vm~~$ & $N\mm~~$ & $\Rm~~$ & $\Vm~~$ & $N\mm~~$ & $\Rm~~$ & $\Vm~~$ & $N\mm~$ \\
\hline
 66.688 & 318.720 &  314996 &  73.758 & 334.129 & 382892 & 10.132 & 150.037 & 10606\\
131.483 & 319.988 &  626008 &  57.948 & 207.773 & 116320 & 12.792 & 150.374 & 13450\\
132.349 & 378.037 &  879489 &  55.595 & 312.906 & 253107 & 17.541 & 131.280 & 14057\\
244.115 & 310.677 & 1095606 &  57.758 & 288.778 & 223965 & 16.190 & 177.786 & 23795\\
178.092 & 368.667 & 1125522 & 159.994 & 354.623 & 935579 & 11.541 & 155.033 & 12898\\
147.816 & 408.469 & 1146778 & 136.115 & 346.092 & 758106 & 13.417 & 218.599 & 29812\\
270.148 & 672.172 &   89802 & 323.188 & 448.259 &  47779 & 20.734 & 209.447 &   669\\
205.886 & 799.660 &   96864 & 140.226 & 647.000 &  43187 & 33.343 & 494.196 &  5991\\
557.573 & 651.487 &  174115 & 246.990 & 486.972 &  43093 & 18.622 & 295.171 &  1193\\
316.831 & 906.057 &  191365 & 349.004 & 802.286 & 165276 & 19.738 & 365.152 &  1936\\
396.234 & 828.811 &  200257 & 219.283 & 591.874 &  56518 & 22.820 & 298.968 &  1500\\
632.925 & 671.631 &  210057 & 206.564 & 484.902 &  35734 & 44.054 & 206.847 &  1386\\
300.471 & 1016.55 &  228448 & 349.004 & 794.472 & 162073 & 77.824 & 604.260 & 20906\\
315.793 & 975.451 &  221074 & 348.943 & 997.308 & 255351 & 26.809 & 356.614 &  2508\\
\hline
\end{tabular}
}
\caption
{
Radius $\Rm$ (in$\hkpc$) where the circular velocity $\Vm$ (expressed in~km~s$^{-1}$) reaches its maximum, and number of particles contained within $\Rm$, for our sample of simulated dark matter haloes, at the three redshifts considered.
}
\label{tabSample}
\end{table*}

For each halo, we compute the position of the centre of mass iteratively, starting with an initial guess and a sphere of $30-300\hkpc$ radius.
When convergence is reached, the radius of the sphere is reduced by ten per cent, and the process is repeated until it encloses less than $10^3$ particles.
This procedure is almost equivalent to selecting the minimum of the gravitational potential as the origin of coordinates \citep[see e.g.][]{Neto+_07}.
Then, $\Vm$ and $\Rm$ are obtained from the circular velocity profile around that point.
Radial bins are chosen so that the enclosed mass increases by 5 per cent (or 100 particles, whichever is larger).
\Revised{
The innermost bin in all our profiles encloses more than 100 particles.
}

To locate the first peak in $\Vc$, we examine the profile in order of increasing radius until the circular velocity decreases below 97.7 per cent of the maximum value found so far; in other words,
\be
\log(\Vc)=0.99\log(\Vm).
\label{eqMax}
\ee
This stopping condition sets the value of $\Vm$.
For the radius $\Rm$, we adopt the geometric average of the two radii fulfilling equation~(\ref{eqMax}).
The origin of velocities is taken as the average velocity of all particles within $0.1\Rm$.
Basic properties of our haloes are summarized in Table~\ref{tabSample}.

The choice of $\Rm$ (and $\Vm$) has several advantages with respect to other alternatives, such as the virial radius or the radius $r_{-2}$ at which the logarithmic slope of the density profile is equal to $-2$.
It is well known that many properties of dark matter haloes (e.g. the density profile) are not universal in terms of the virial radius only.
Moreover, this quantity is not well defined for substructures, and tends to be unstable even for haloes undergoing weak interactions with close neighbours.
While this does not pose a problem for the study of relaxed, isolated systems, it is not the optimal choice in the general case.
The radius $r_{-2}$, on the other hand, provides a much better reference point in the sense that many profiles display a universal structure when scaled in units of $r_{-2}$, and it is much less sensitive to the environment than the virial radius.
However, it has the technical disadvantage that it is defined in terms of the local slope of the density profile, which is a second derivative of the mass distribution and therefore prone to significant numerical errors.
This difficulty can be overcome by assuming an analytical model of the dark matter halo.
In this case, $\Rm$ and $r_{-2}$ are proportional to one another, and the choice between them becomes a matter of personal preference.
We opted for $\Rm$ because this quantity is model-independent; in fact, it does not rely on the assumption of a universal shape of the density profile.
Moreover, the circular velocity of the halo is closer to an observable quantity than the halo mass, let alone the logarithmic slope of the density, and it reflects the depth of the potential well of the whole halo rather than a local property, which makes it a more robust estimator of the actual scale of the object.

  \section{Results}
  \label{secResults}

In addition to the mass and circular velocity profiles, we obtained the dark matter density profile and its logarithmic slope.
The average radial velocity and the global angular momentum are also computed within spherical shells, as well as the radial and tangential components of the velocity dispersion.
This makes possible to investigate the radial dependence of other physical properties, such as the anisotropy parameter or the coarse-grained phase-space density.

For all these quantities, we first consider the existence of a universal profile in terms of the two fundamental parameters of the halo, which we take to be $\Rm$ and $\Vm$.
When appropriate, we compare our numerical results with the analytical models proposed by \citet[hereafter NFW]{NFW97}, \citet[hereafter R04]{Rasia+04} and \citet[hereafter DM05]{DehnenMcLaughlin05}.
In the NFW model, the maximum circular velocity occurs at $\Rm\simeq2.163\,r_{-2}$, while in R04, it takes place at $\Rm\simeq0.334\,R\vir$.
For the DM05 model, we assume the quoted canonical values, $\epsilon=3$, $\beta_0=-0.1$ and $\beta_\infty=0.67$.
For this particular choice of the parameters, the maximum circular velocity occurs at a radius $\Rm\simeq1.852\,r_0$, where $r_0$ denotes a characteristic radius of the halo.

Possible systematic dependences on halo mass are investigated by considering the results from our two sets of simulations (containing galaxy- and cluster-size objects at $z=0$) separately.
Temporal evolution is addressed by analyzing the outputs at $z=0$, 1, and 5.

\subsection{Mass distribution}

\begin{figure*}
\includegraphics[width=5.5cm]{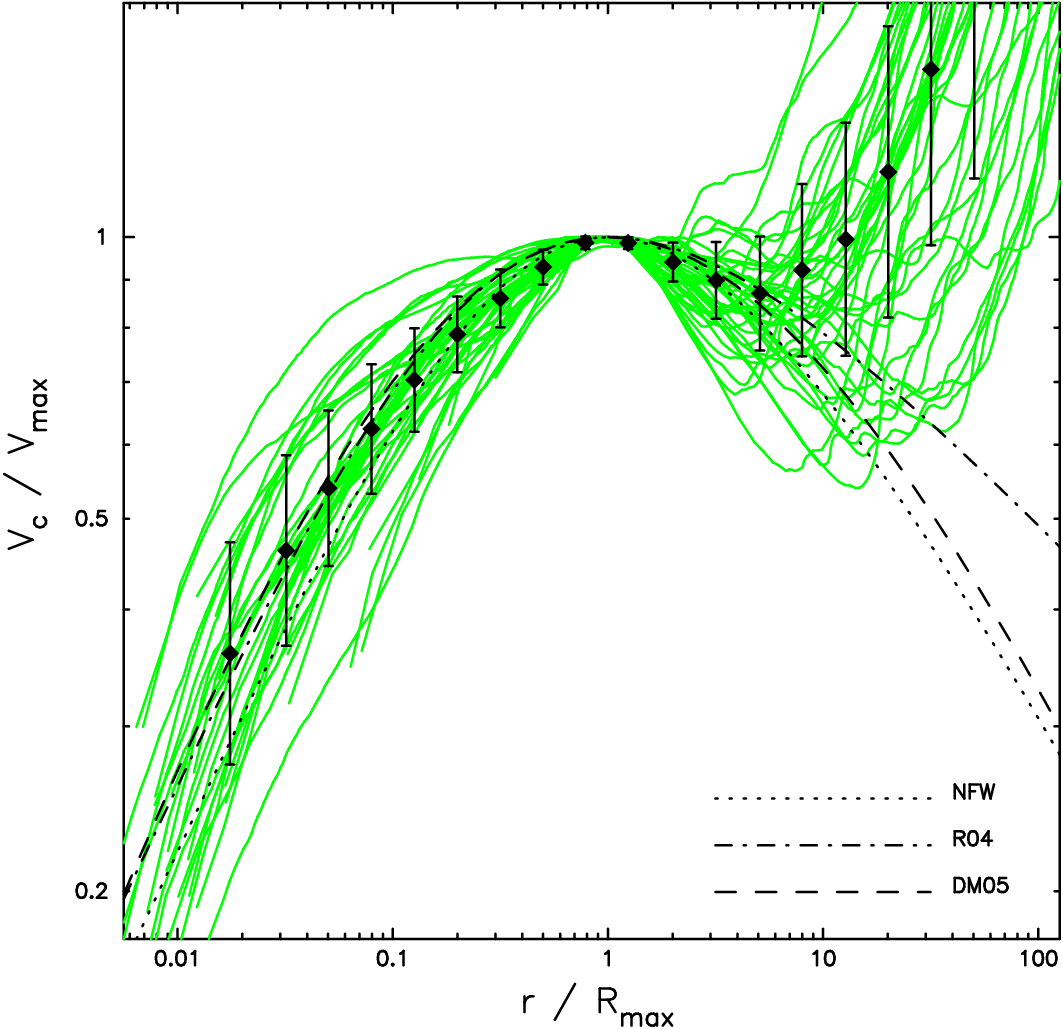}
\hfill
\includegraphics[width=5.5cm]{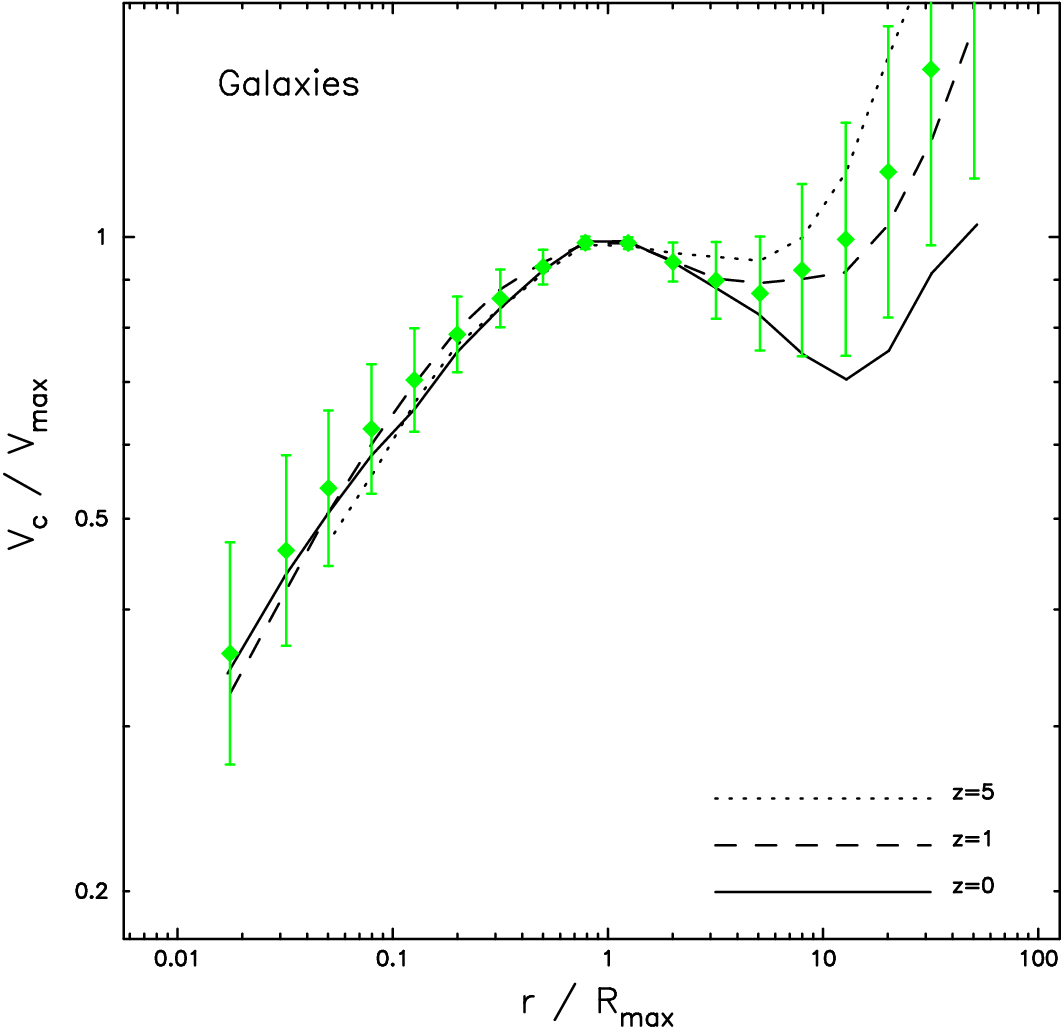}
\hfill
\includegraphics[width=5.5cm]{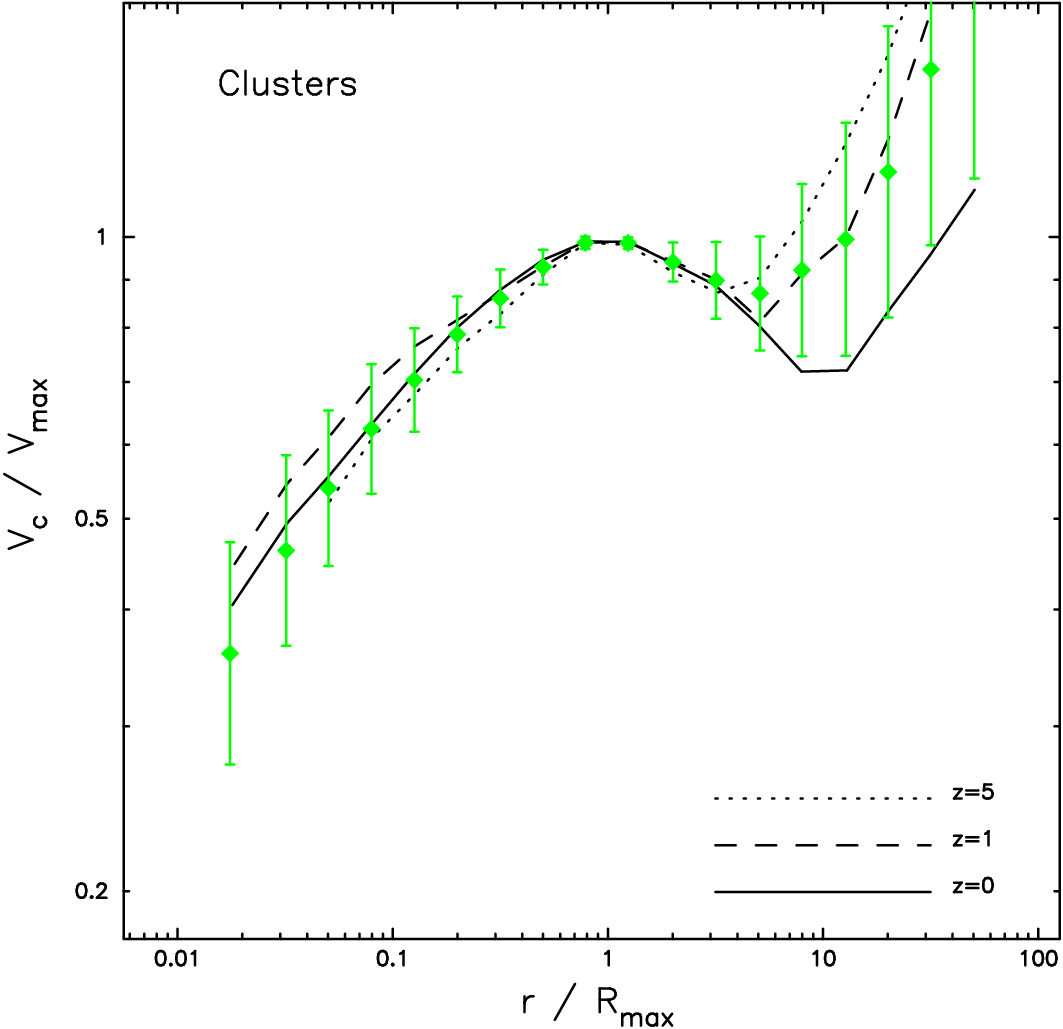}
\caption
{
  Circular velocity profiles, $\Vc=\sqrt{GM/r}$, of our sample of dark matter haloes.
  Points with error bars show the average and one-$\sigma$ scatter, including all objects, at all redshifts.
  On the left panel, individual profiles are plotted as green solid lines; dotted, dot-dashed, and dashed lines represent the NFW, R04 and DM05 models, respectively.
  On the middle and right panels, the global average (shown in green) is compared to the results obtained from our galaxy and cluster simulations.
  In both panels, solid, dashed and dotted lines correspond to the average profile at $z=0$, 1, and 5, respectively.
}
\label{figVc}
\end{figure*}
\begin{figure*}
\includegraphics[width=5.5cm]{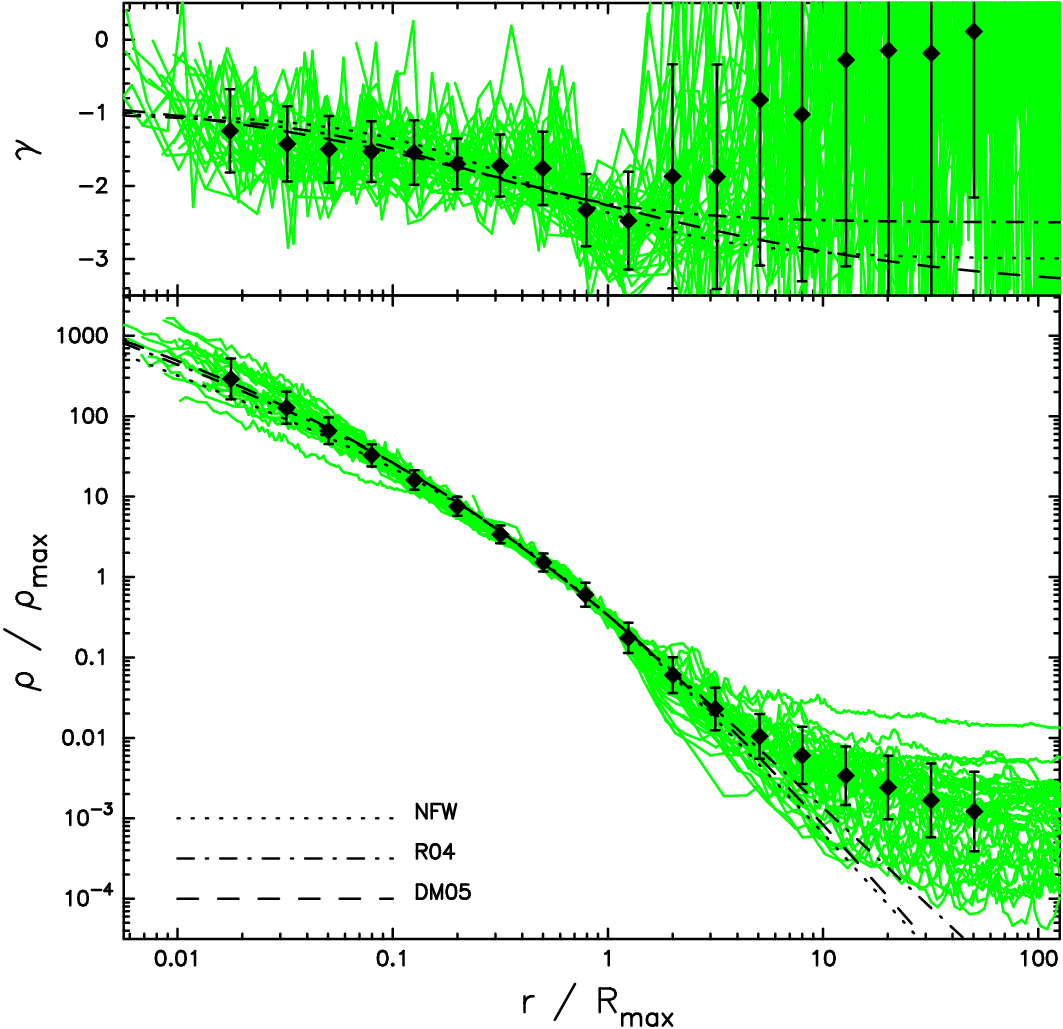}
\hfill
\includegraphics[width=5.5cm]{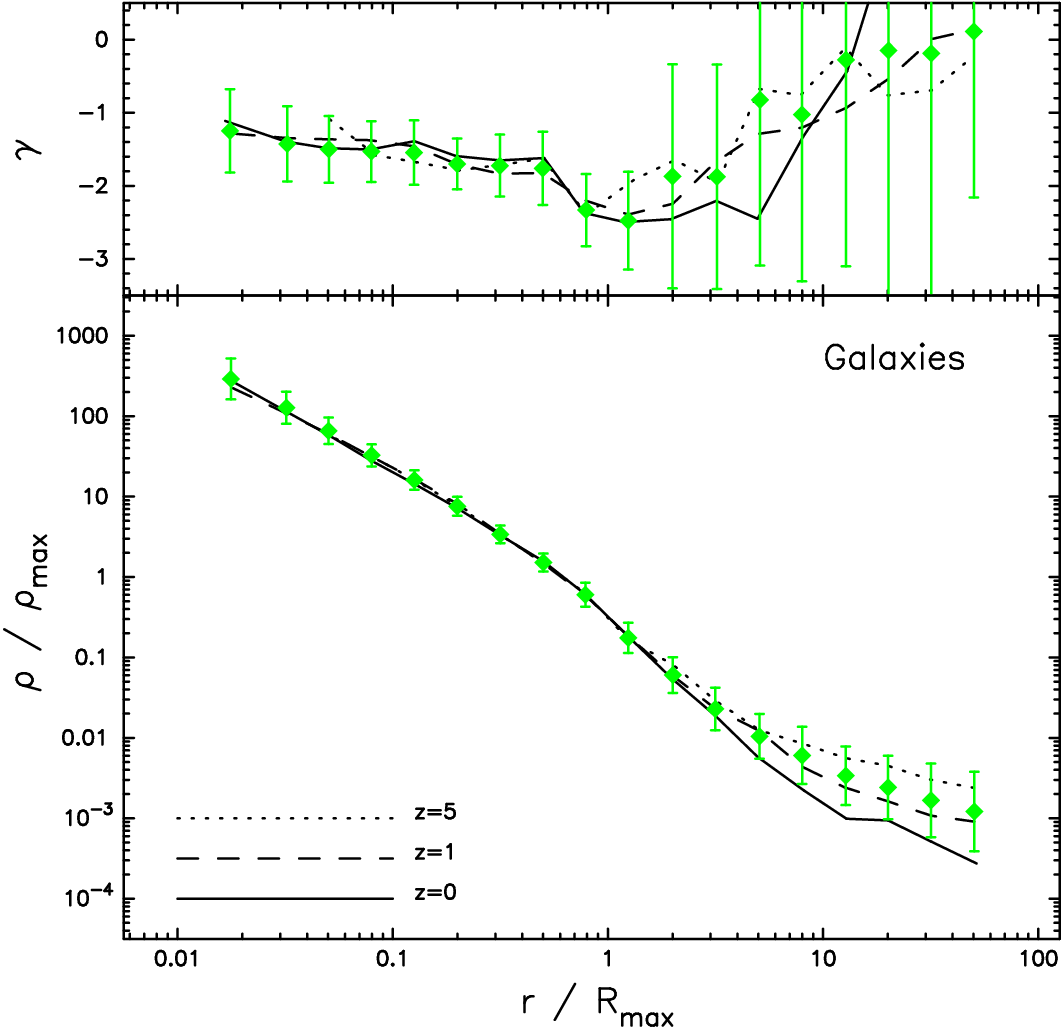}
\hfill
\includegraphics[width=5.5cm]{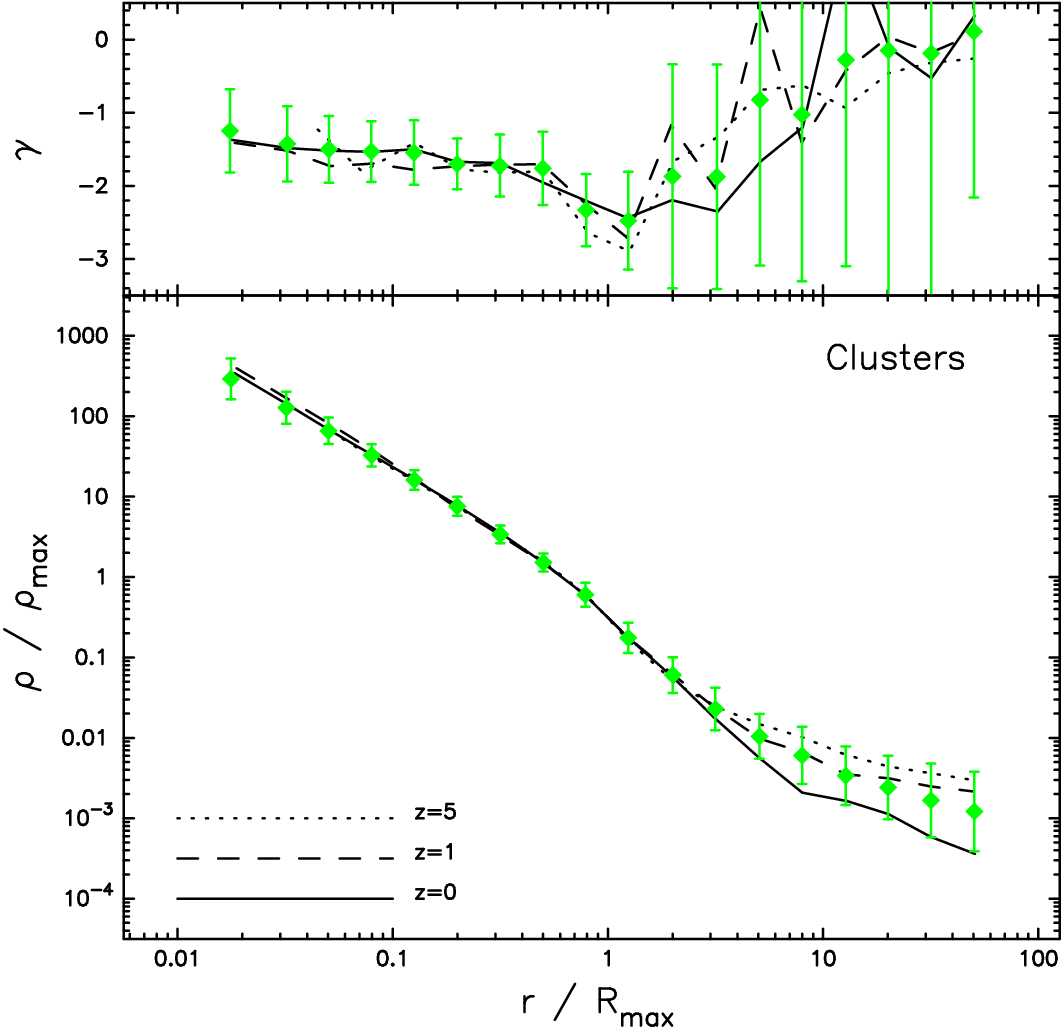}
\caption
{
  Density profiles (bottom) and their logarithmic slope (top), scaled by the characteristic density $\rho\mm$ and radius $\Rm$ of each halo.
  Line styles are the same as in Figure~\ref{figVc}.
}
\label{figRho}
\end{figure*}

The circular velocity profiles of our 14 dark matter haloes, at the three different redshifts under study, are plotted together on the left panel of Figure~\ref{figVc}.
Each individual profile is normalized at its maximum value, $\Vm$, and the radius at which it is attained, $\Rm$.
All haloes seem to be well described by a universal circular velocity profile; the scatter around the average, indicated by the error bars in Figure~\ref{figVc}, is smaller than 30 per cent at all radii.
This result, though not surprising, is remarkable in the sense that $\Vm$ and $\Rm$ are not fits to the dark matter distribution, but physical properties of the halo.
Indeed, the radius $\Rm$ (as well as $r_{-2}$) corresponds to the transition between two completely different dynamical regimes\footnote{For $r<\Rm$, the mass is dominated by particles with apocentre $a>r$ (i.e. coming from outer regions). Outside $\Rm$, most particles have $a<r$ \citep[see e.g.][]{FillmoreGoldreich84,Bertschinger85}.}, and, particularly at high redshift, we claim it is a much better indicator of the region within which the halo can be considered in equilibrium.
A first hint into that direction can already be seen on the middle and right panels, which show the evolution of galaxies and (proto)clusters, respectively.
Whereas cosmic evolution plays an important role outside $\Rm$, the average behavior of the circular velocity profile is much more homogeneous within this radius.

A similar conclusion can be reached in terms of the density profile.
The density profiles of our dark matter haloes are plotted in Figure~\ref{figRho}, normalized to the characteristic density
\be
\rho\mm = \frac{3}{4\pi G} \left( \frac{\Vm}{\Rm} \right)^2
\ee
and the radius $\Rm$.
Again, neither the density profile nor its logarithmic slope $\gamma$ (plotted on the top sub-panels) deviate significantly from the average behaviour for $r<\Rm$.
In the outer regions, one does not only observe the effects of cosmic evolution, but also a significant variation from object to object (much larger than any dependence on halo mass) which reflects the different environments in which our dark matter haloes are embedded.
Within $\Rm$, we do not observe any systematic trend with mass or redshift, but one should bear in mind that small departures from universality (in particular, smaller than the scatter) could only be detected with a much larger sample of simulated objects.

Concerning the precise shape of the density or circular velocity profiles, our results are consistent up to $\Rm$ with all of the three analytical models considered.
The average density profile of our haloes seems to be somewhat steeper than the NFW formula, and it is better reproduced by the functions proposed by R04 or DM05.
This, however, is probably related to the fact that we have not made any selection based on the dynamical state of our haloes, while the NFW model is intended to provide a good description of relaxed objects (see e.g. the discussion in R04).

\begin{figure*}
\includegraphics[width=5.5cm]{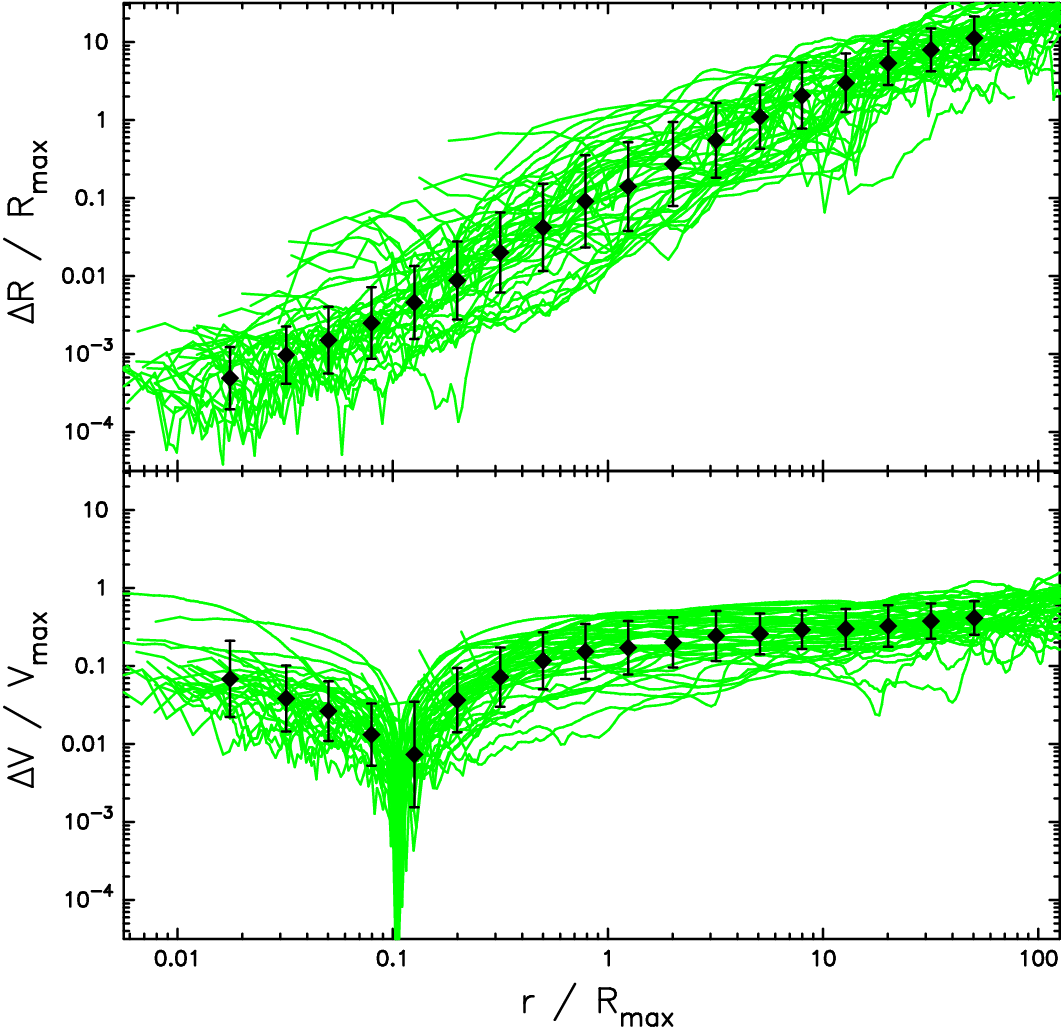}
\hfill
\includegraphics[width=5.5cm]{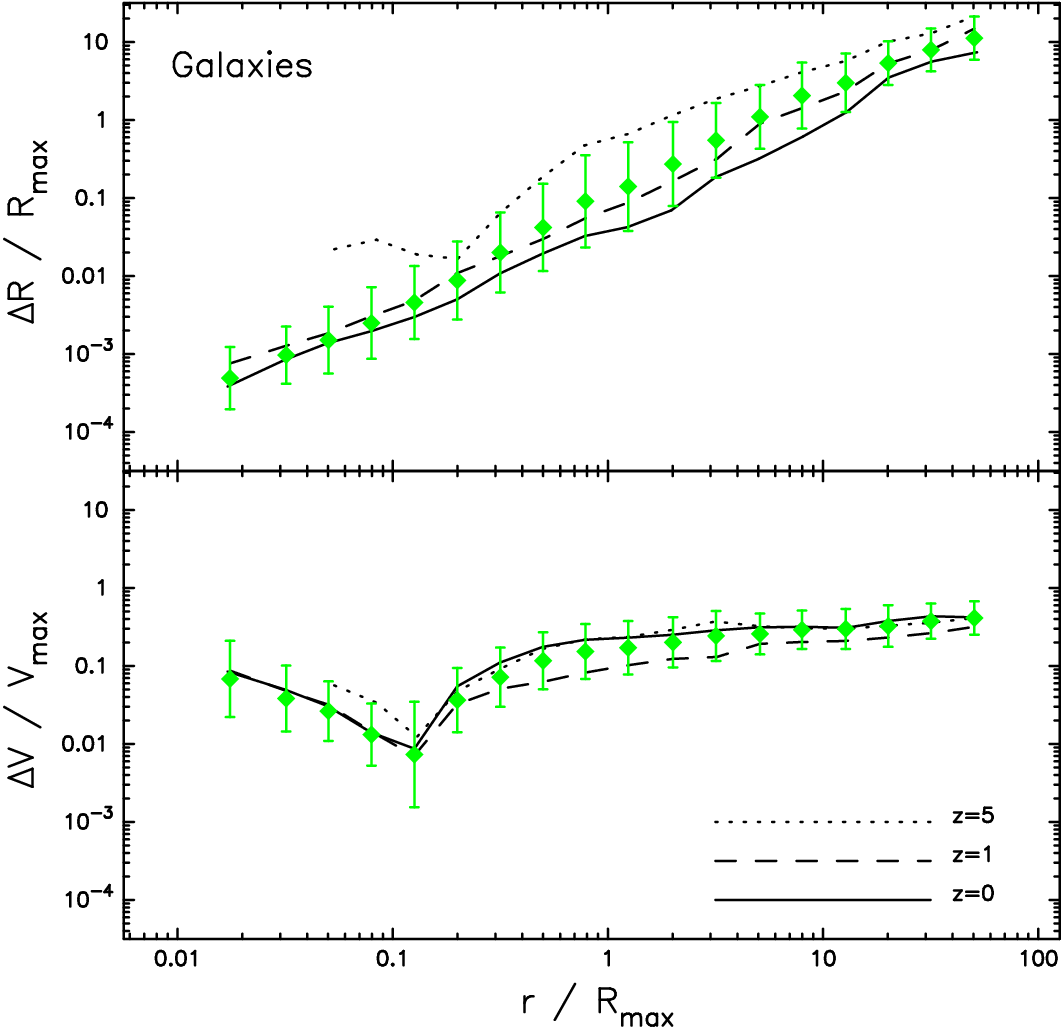}
\hfill
\includegraphics[width=5.5cm]{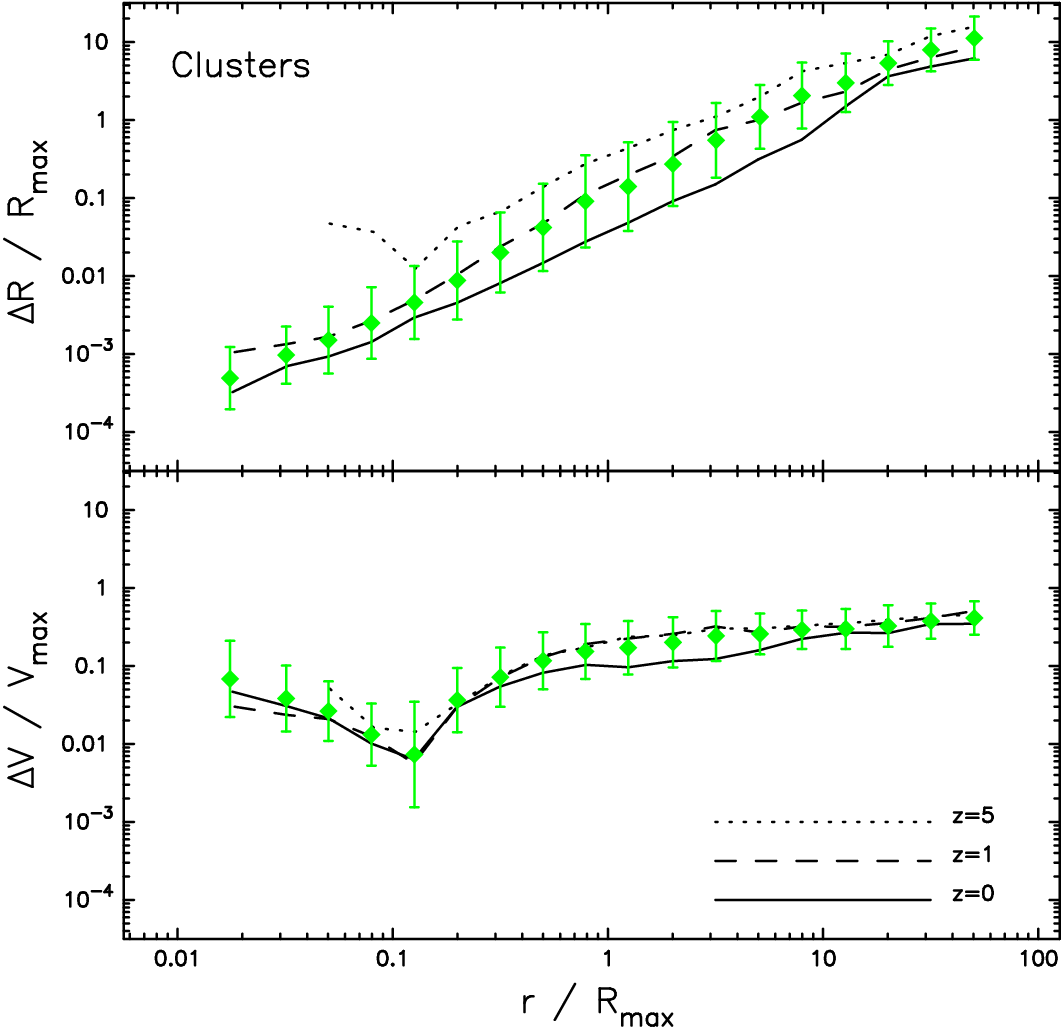}
\caption
{
  Modulus of the offset of the position (top) and velocity (bottom) of the centre of mass within radius $r$ with respect to the central density peak.
  Line styles are the same as in Figure~\ref{figVc}.
}
\label{figOffset}
\end{figure*}
\begin{figure*}
\includegraphics[width=5.5cm]{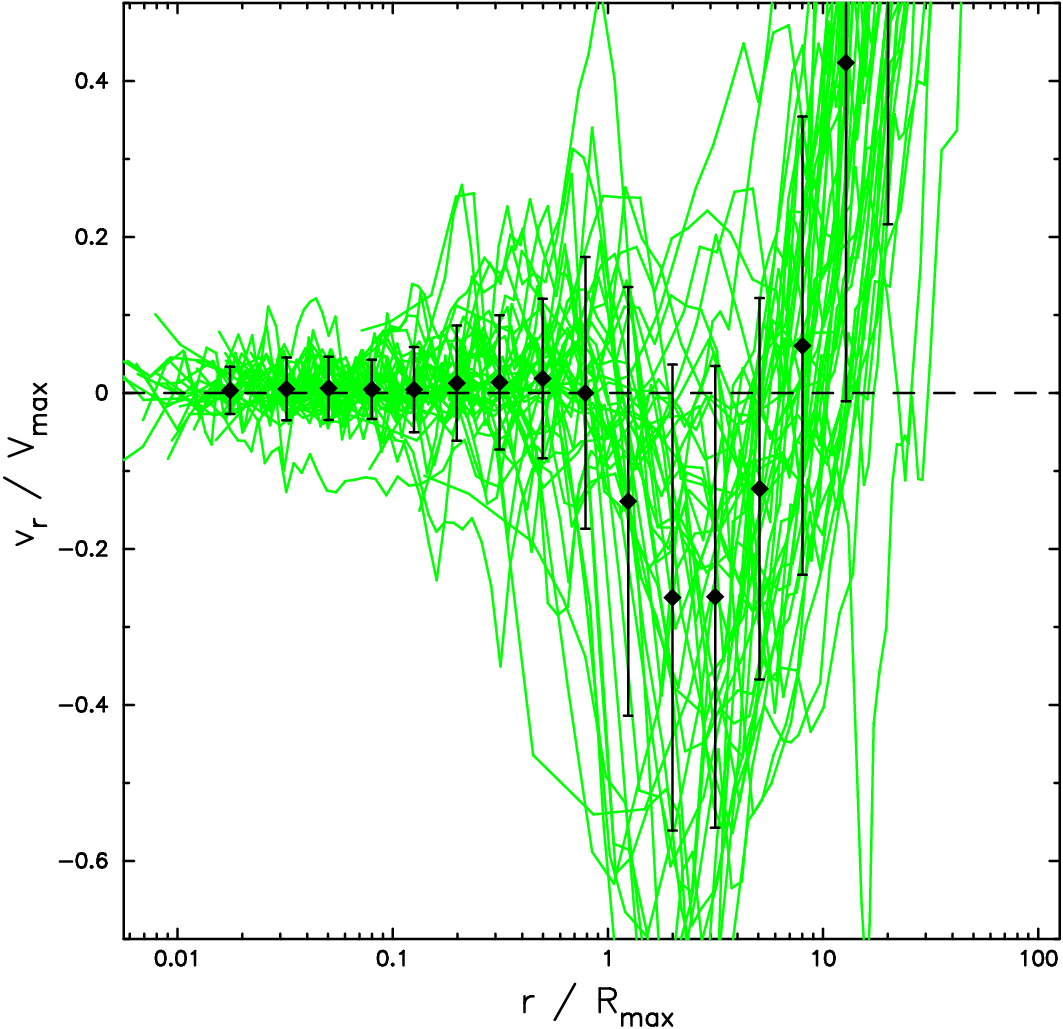}
\hfill
\includegraphics[width=5.5cm]{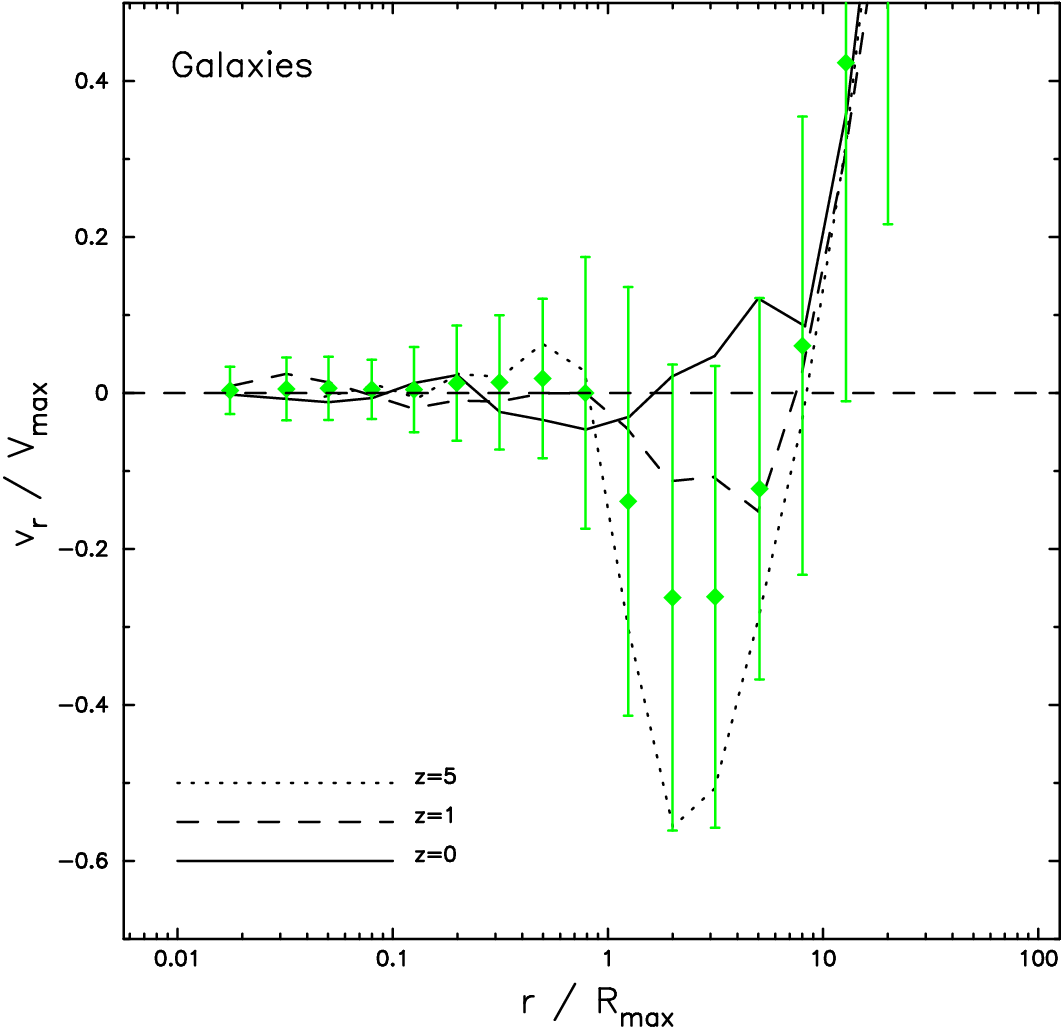}
\hfill
\includegraphics[width=5.5cm]{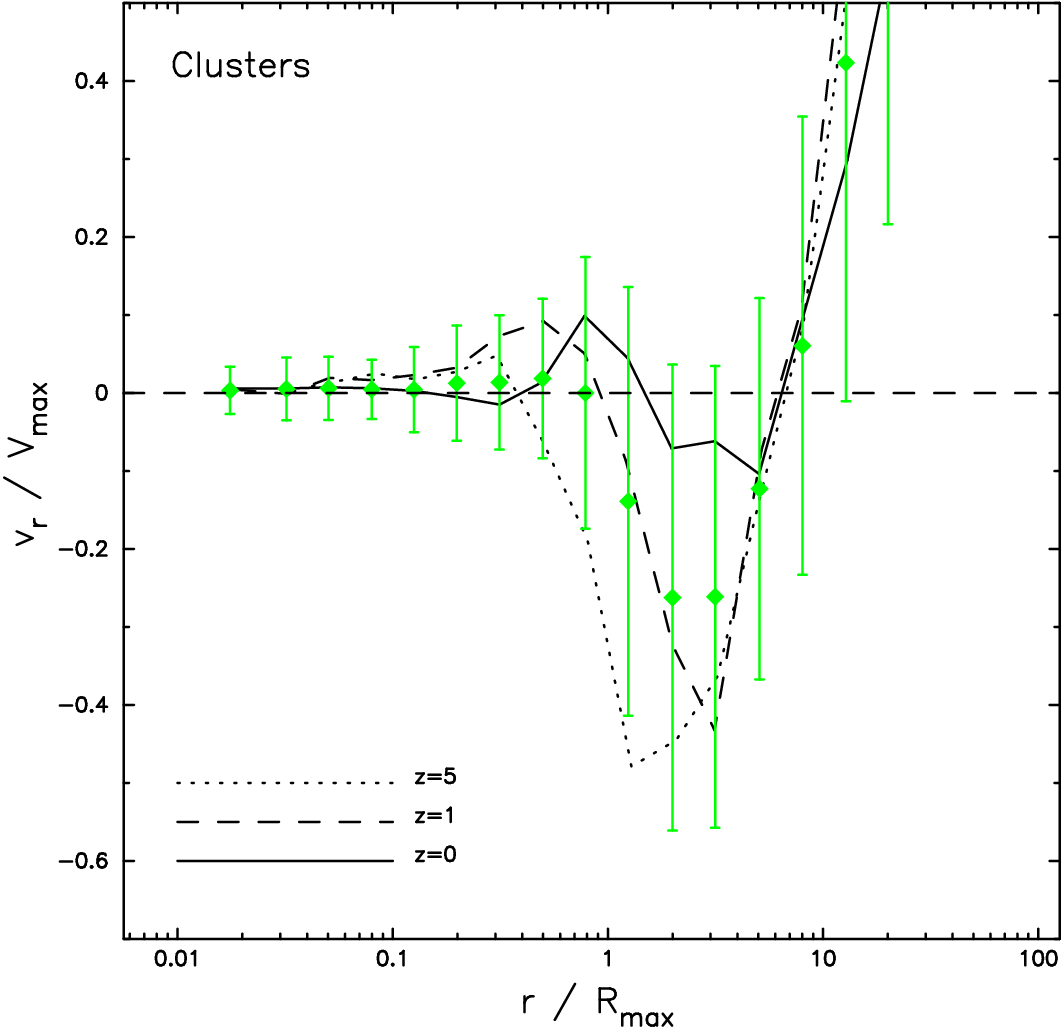}
\caption
{
  Radial velocity profiles, including Hubble flow.
  Line styles are the same as in Figure~\ref{figVc}.
}
\label{figVr}
\end{figure*}

\subsection{Bulk motion}

\begin{figure*}
\includegraphics[width=5.5cm]{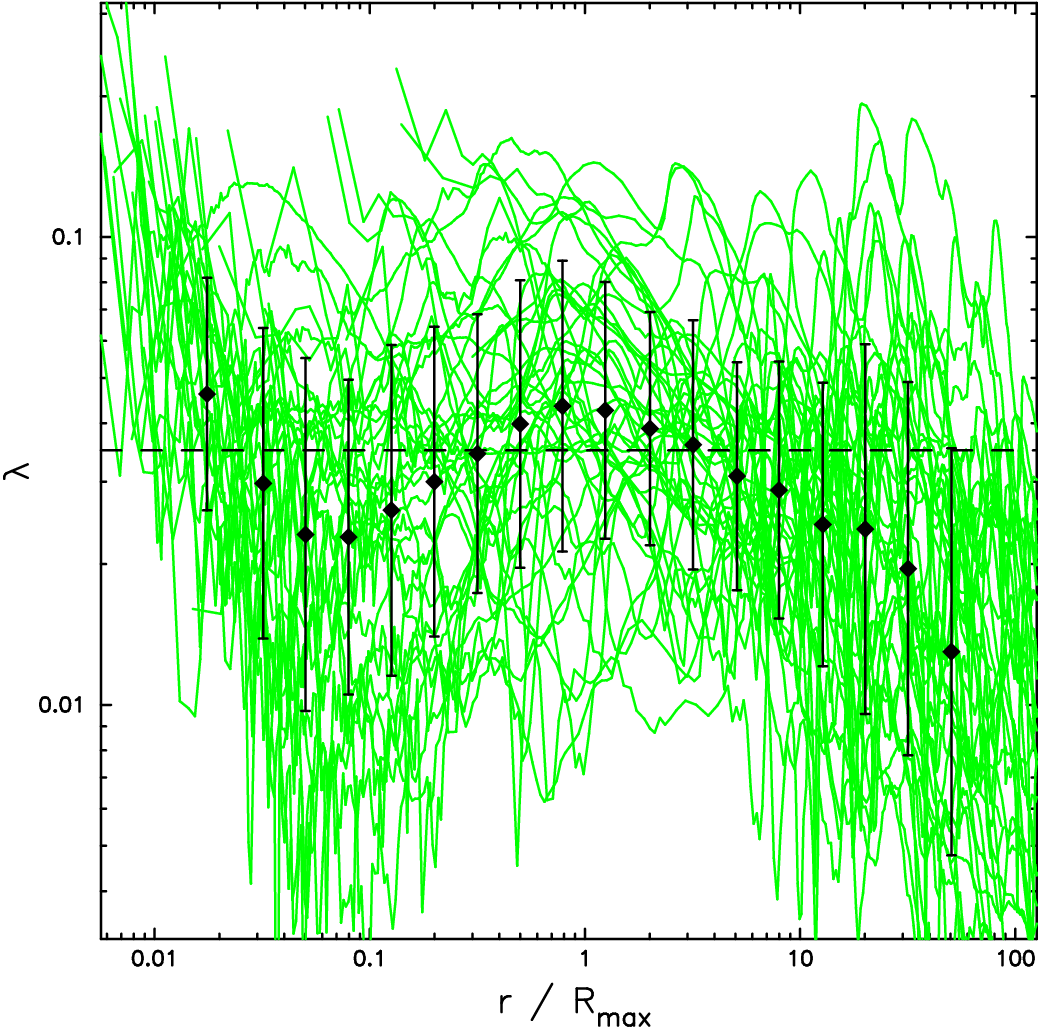}
\hfill
\includegraphics[width=5.5cm]{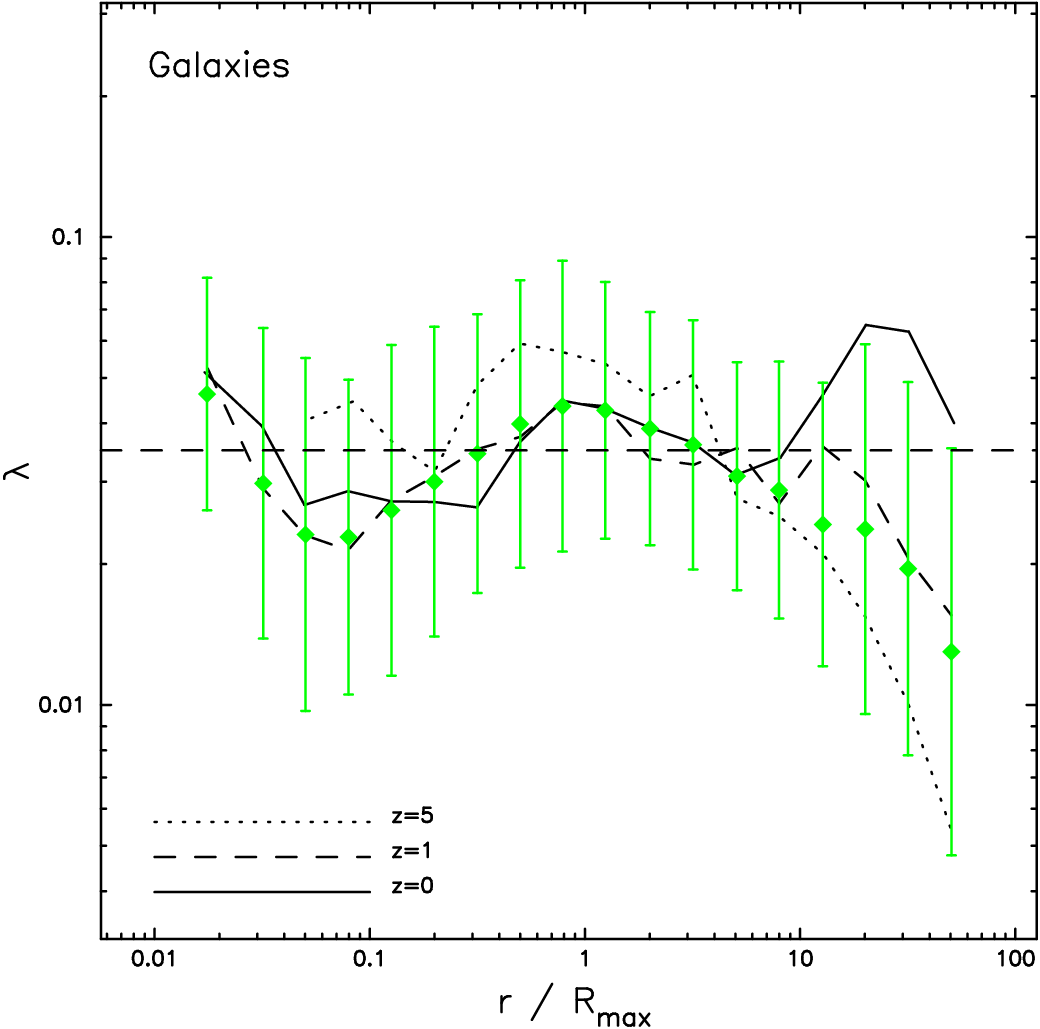}
\hfill
\includegraphics[width=5.5cm]{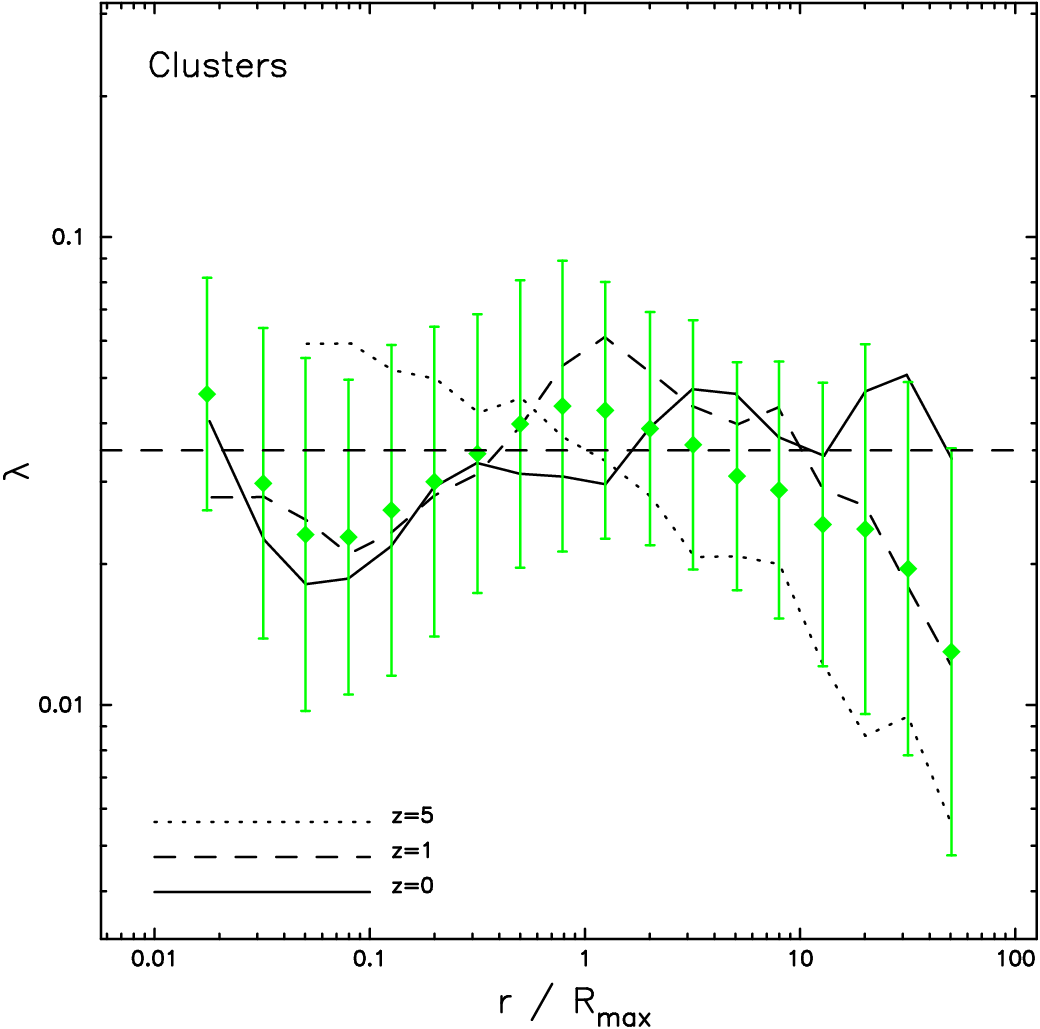}
\caption
{
  Spin parameter, $\lambda=J/\sqrt{2GM^3r}$.
  A horizontal dashed line is drawn at $\lambda=0.035$; other line styles are the same as in Figure~\ref{figVc}.
}
\label{figSpin}
\end{figure*}
\begin{figure*}
\includegraphics[width=5.5cm]{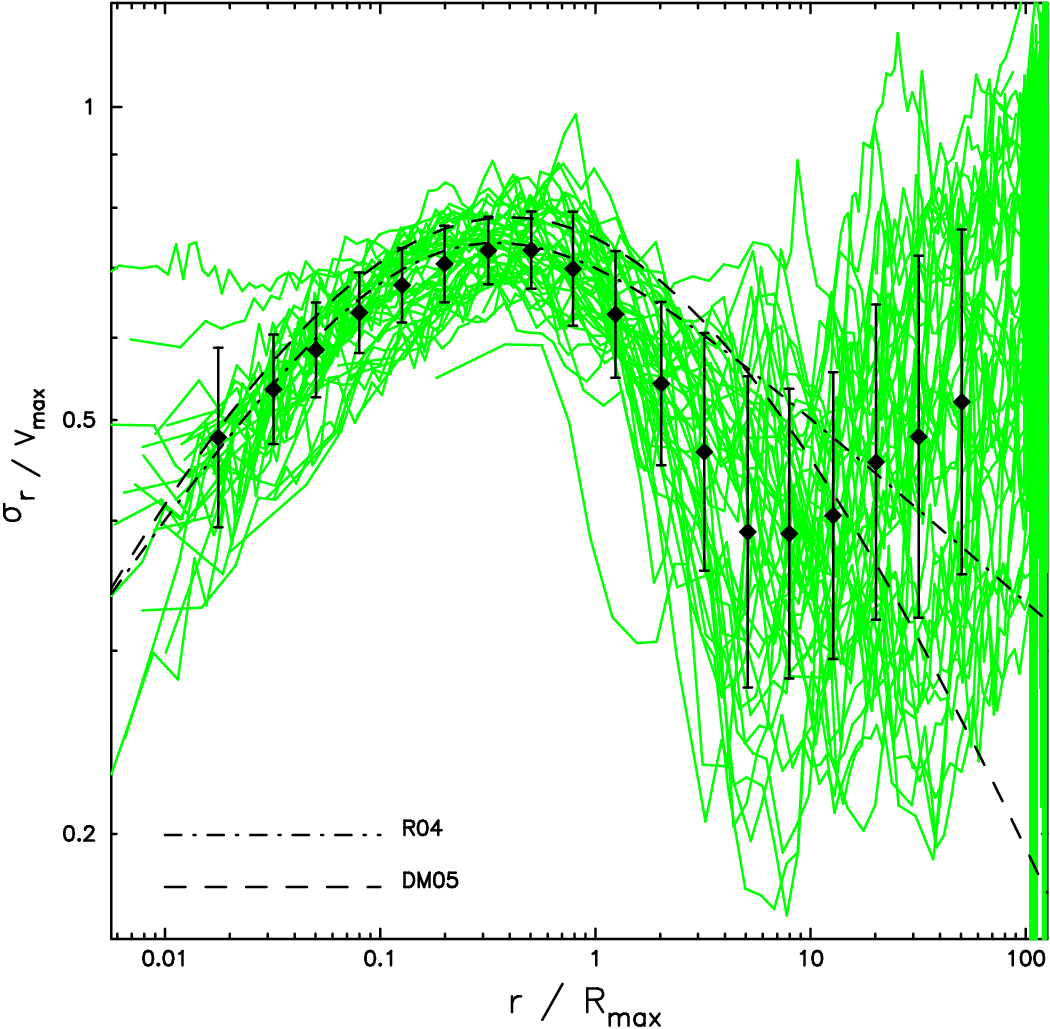}
\hfill
\includegraphics[width=5.5cm]{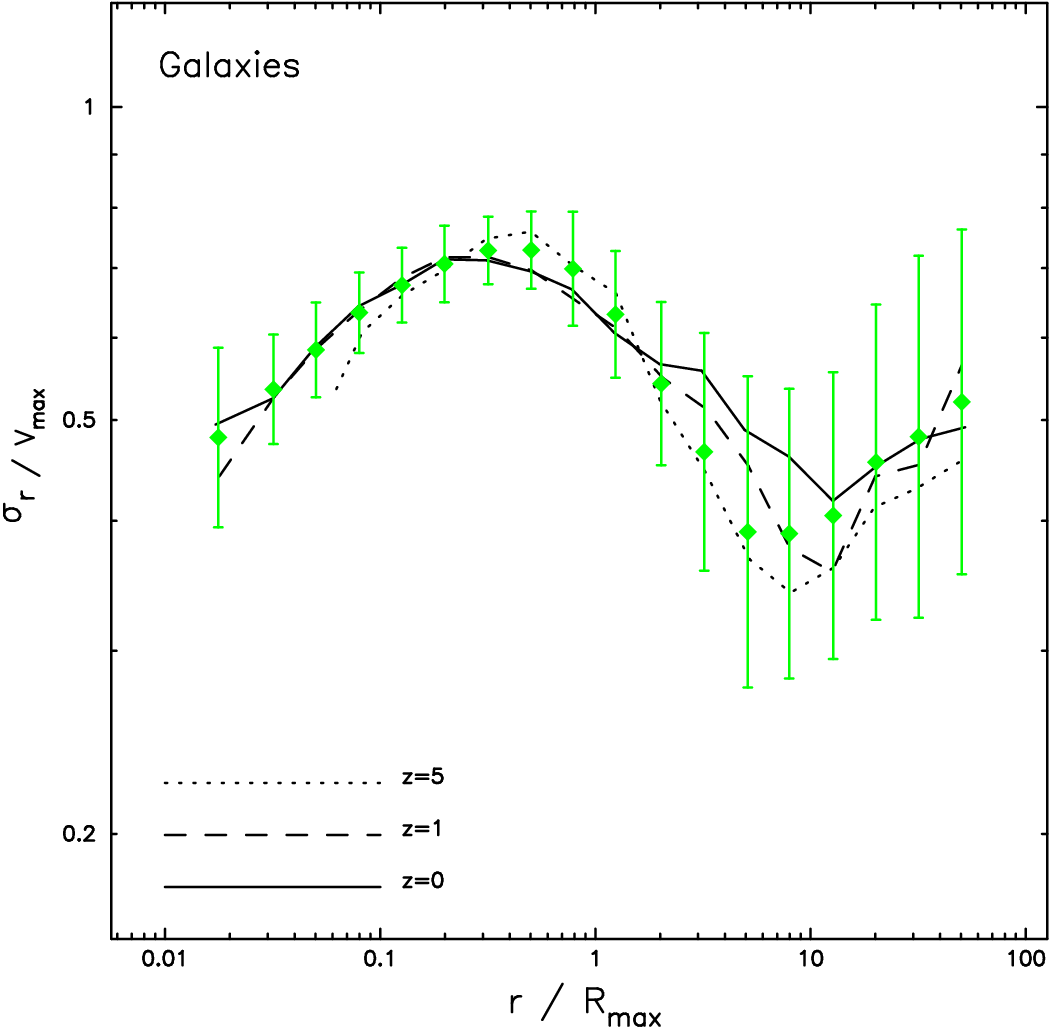}
\hfill
\includegraphics[width=5.5cm]{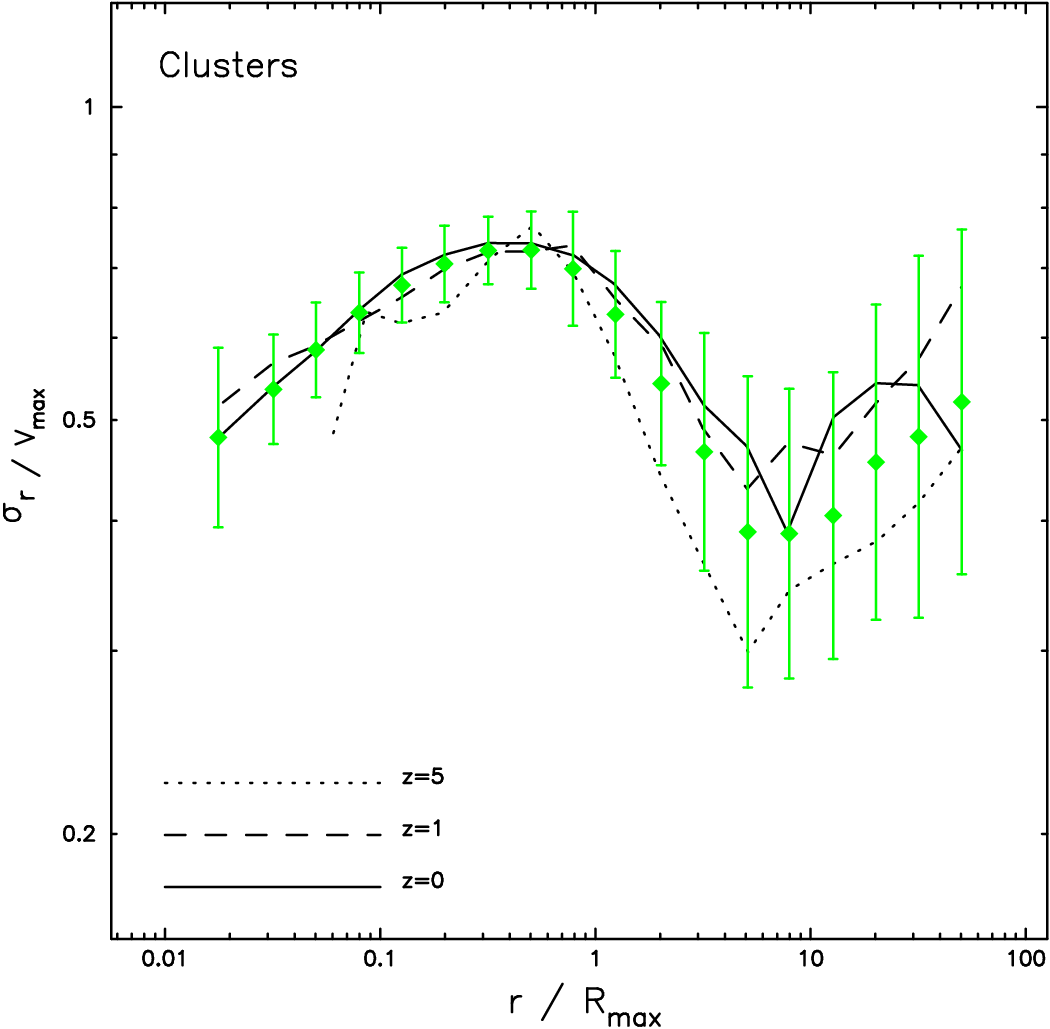}
\caption
{
  Radial velocity dispersion.
  Line styles are the same as in Figure~\ref{figVc}.
}
\label{figSigmar}
\end{figure*}
\begin{figure*}
\includegraphics[width=5.5cm]{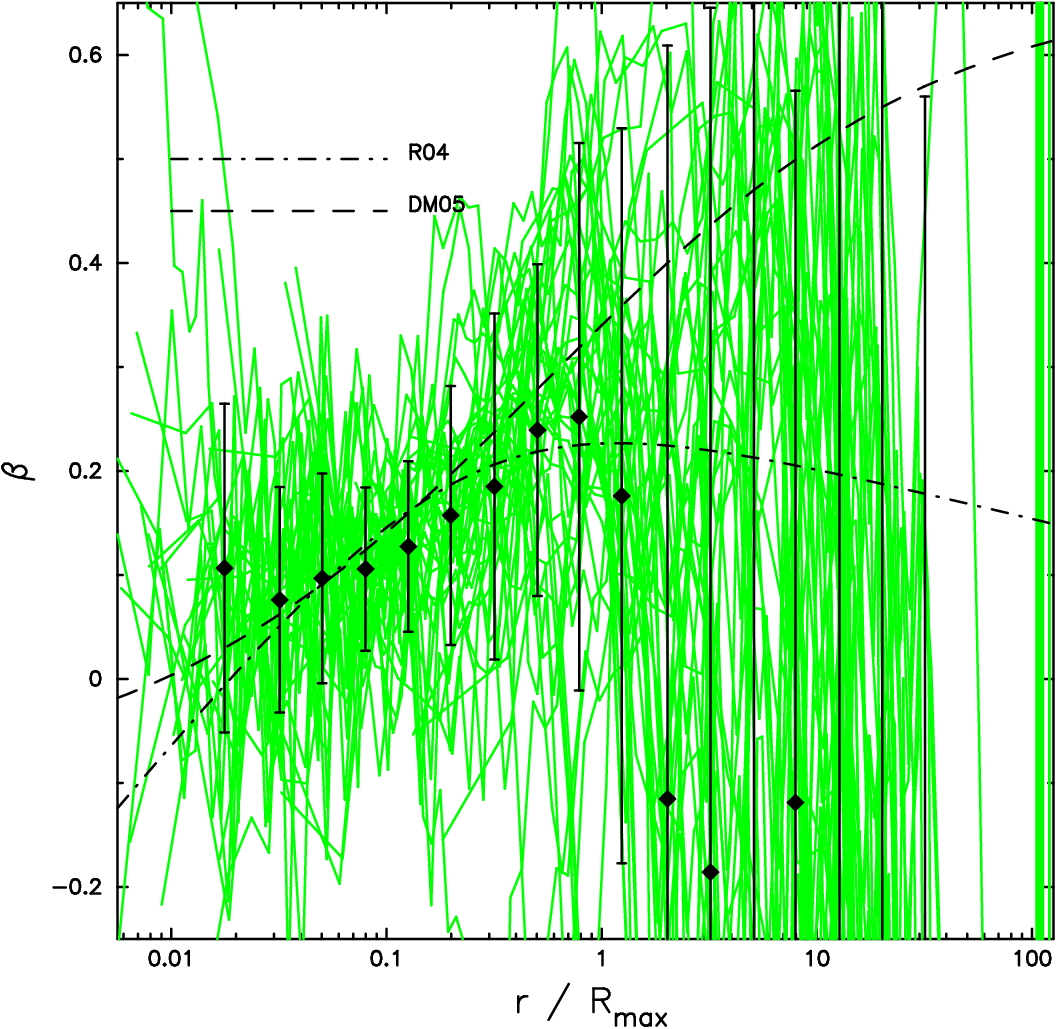}
\hfill
\includegraphics[width=5.5cm]{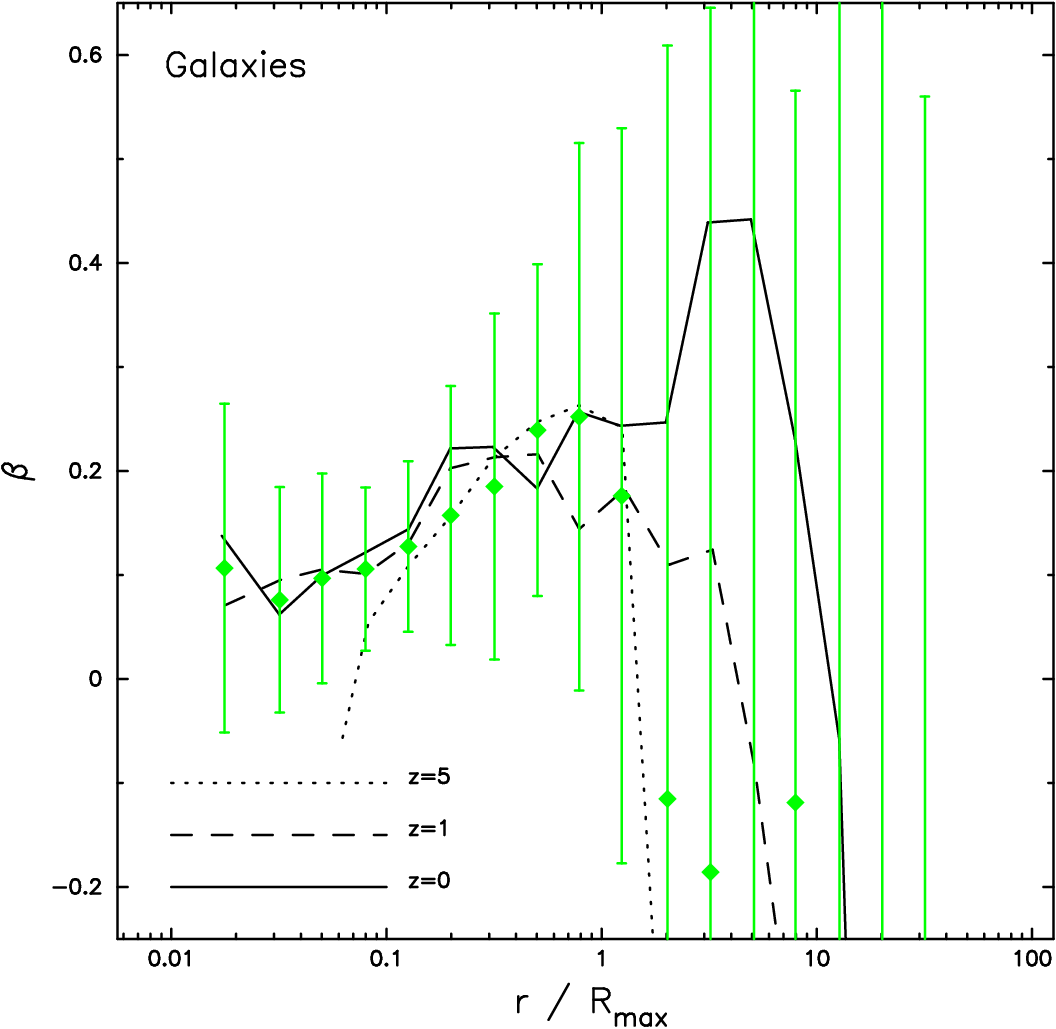}
\hfill
\includegraphics[width=5.5cm]{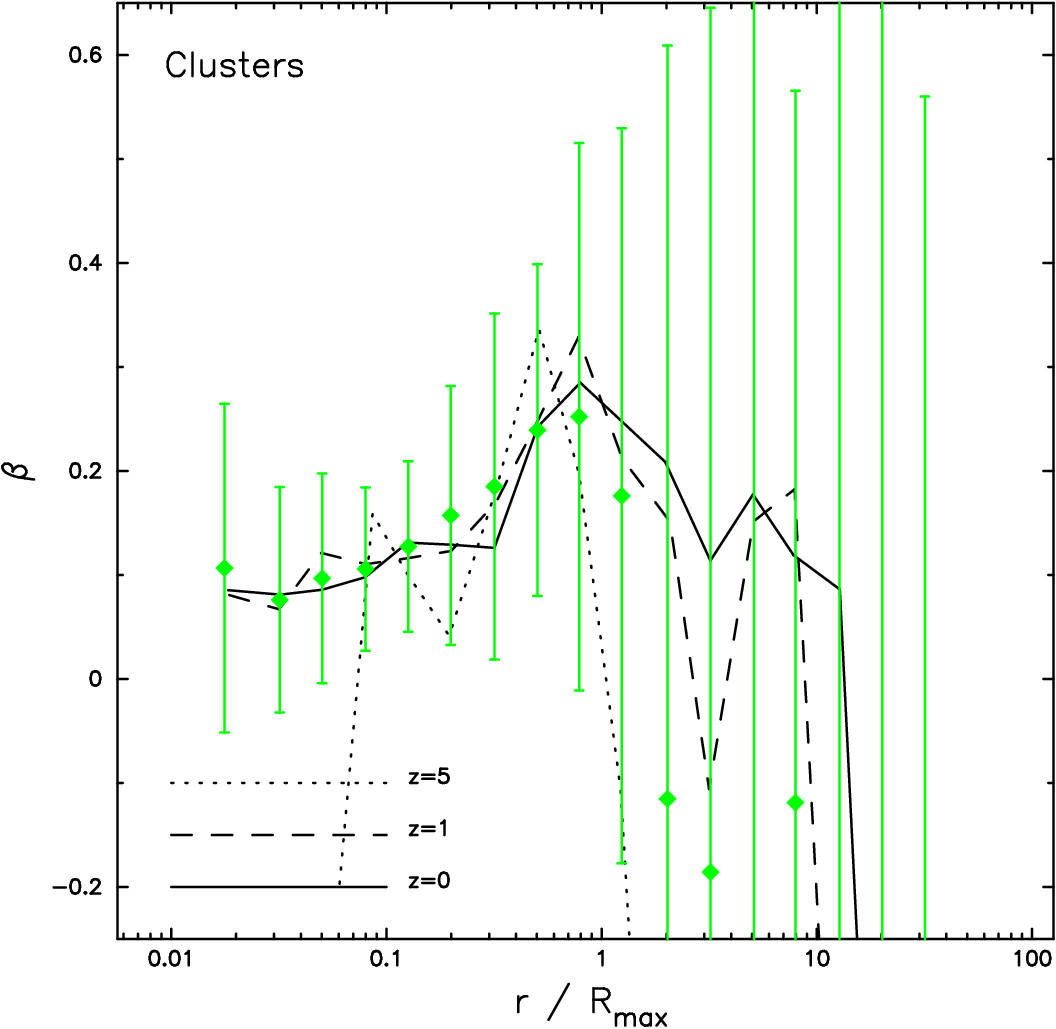}\\[2mm]
\includegraphics[width=5.5cm]{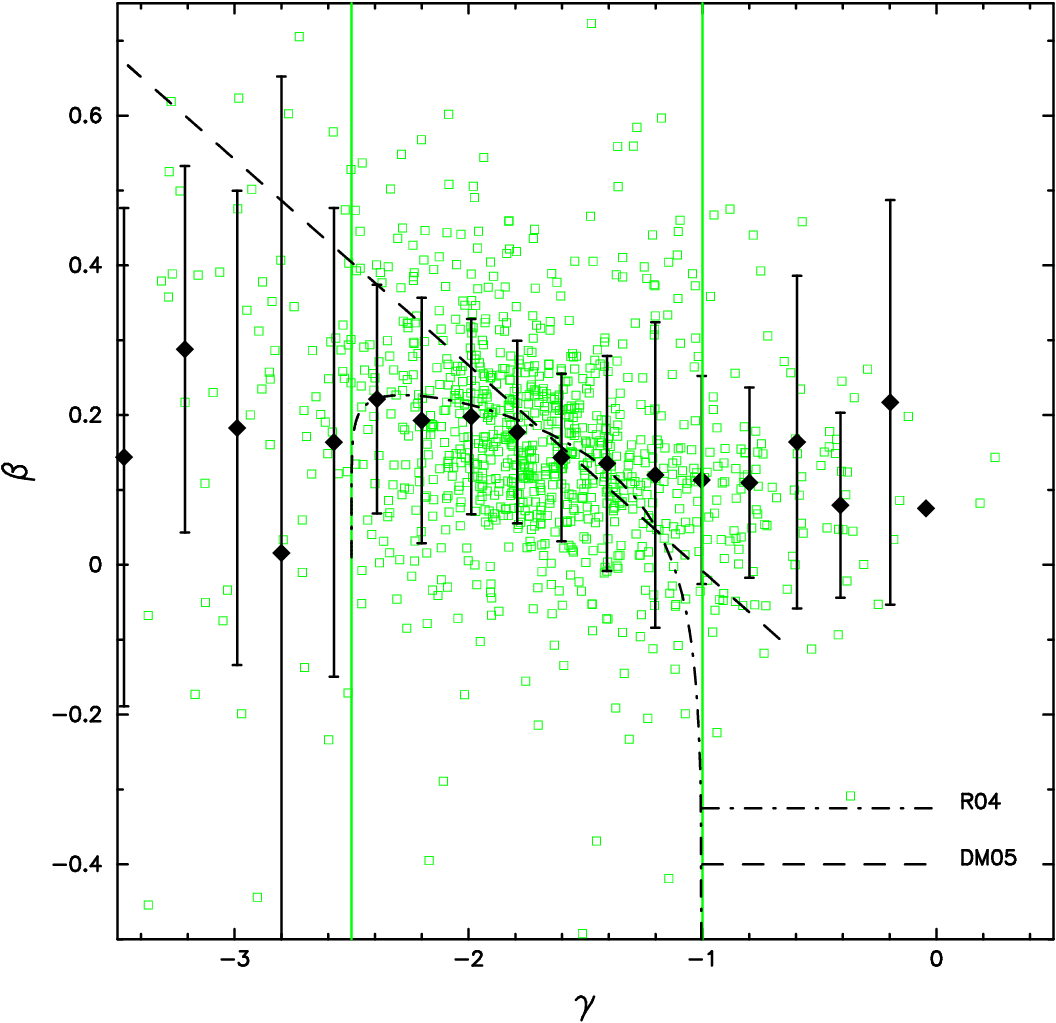}
\hfill
\includegraphics[width=5.5cm]{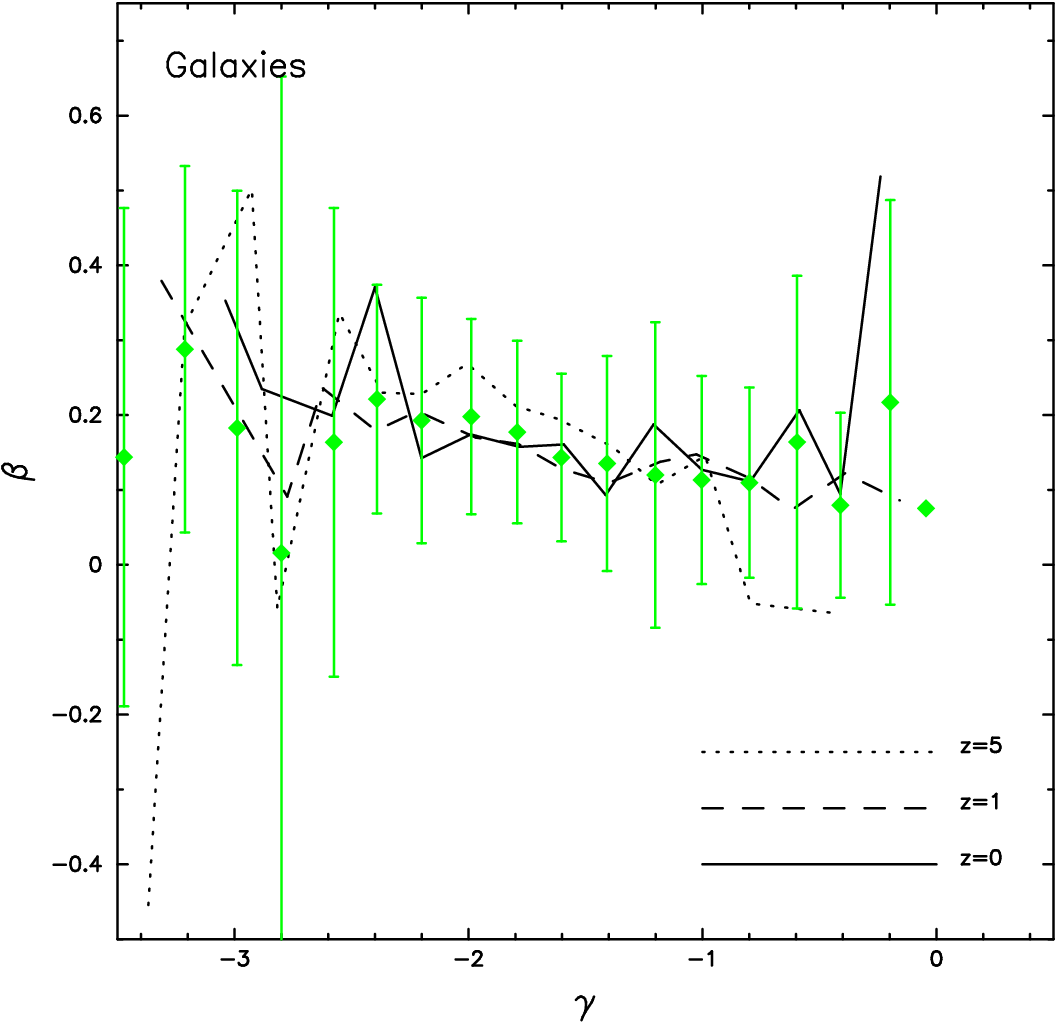}
\hfill
\includegraphics[width=5.5cm]{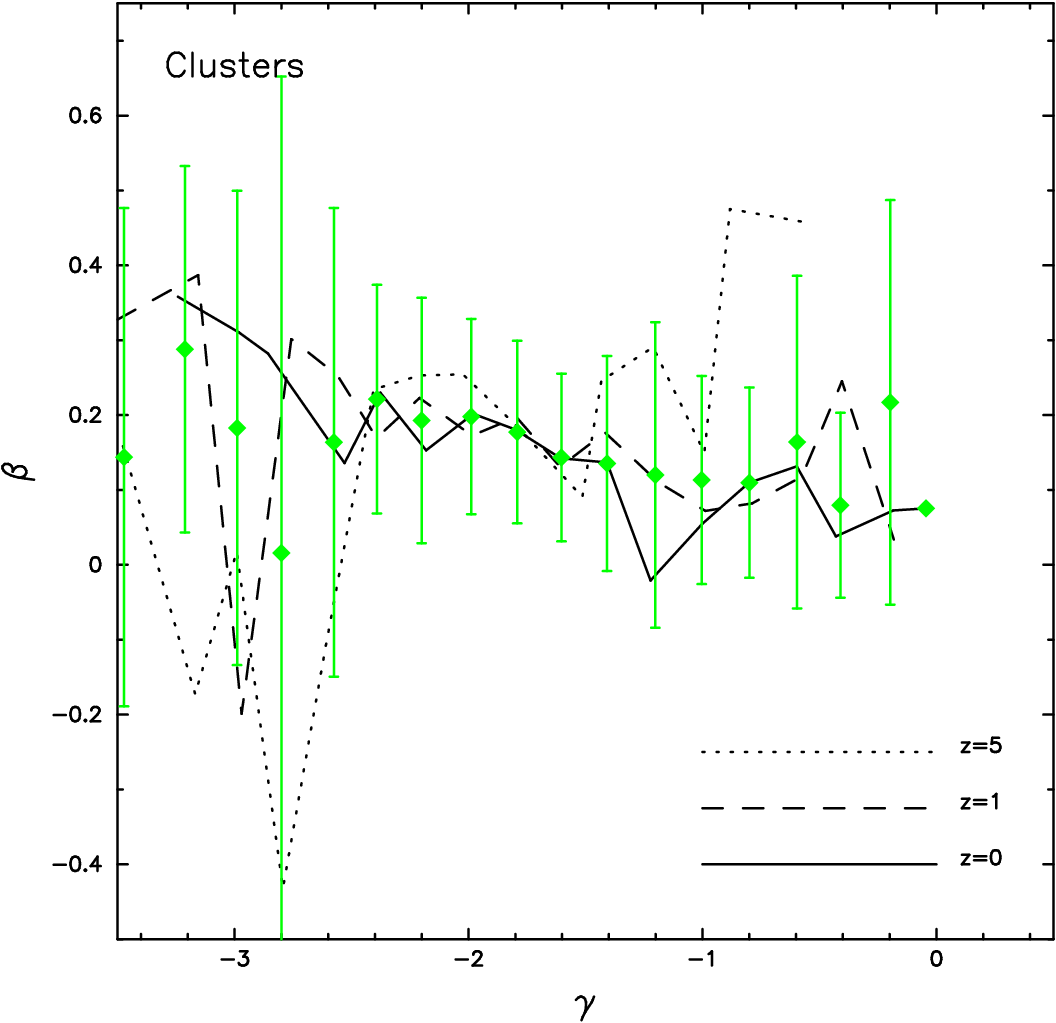}
\caption
{
  Anisotropy parameter (top) and and its relation with the local logarithmic slope of the density profile (bottom).
  Line styles are the same as in Figure~\ref{figVc}.
  \Revised{
    Open squares on the bottom left panel represent individual measurements of $\gamma$ and $\beta$ at a given radius (only data whithin $\Rm$ are considered), and vertical lines indicate the range used to fit the $\beta-\gamma$ relation.
  }
}
\label{figBeta}
\end{figure*}

The average velocity field of dark matter haloes can be separated into a radial and a tangential component; radial motion tells us about the infall of matter into the halo, while tangential motion, measured by the spin parameter, contains information about the recent accretion history and the surrounding tidal field.
On the other hand, the offsets between the coordinates (position and velocity) of the centre of mass on different scales reflect the displacement of the halo with respect to the nearby large-scale structures.

Let us consider first the motion of the dark matter halo as a whole.
If haloes were smooth, spherically-symmetric systems in perfect equilibrium, the position of centre of mass should be independent of the scale considered.
This statement would also be true in triaxial, rotating systems, and it holds even in the case of significant infall, as long as the mass contributed by substructures is small compared to the total mass of the material being accreted.
Therefore, one would expect that a possible offset between the central and outer regions of the halo should be small in relaxed systems.
Along these lines, the distance between the most bound particle (close to our adopted centre) and the centre of mass evaluated at the scale of the virial radius has been used to detect unrelaxed haloes and quantify departures from equilibrium \citep[see e.g.][]{Maccio+07,Neto+_07}.

Figure~\ref{figOffset} shows the offset of the position and the velocity of the centre of mass (evaluated within a sphere of radius $r$) with respect to our adopted origin of coordinates, which traces the highest density peak.
The behaviour of the centre of mass velocity (plotted on the bottom panels) is easy to understand: the offset is minimum at $r=0.1\Rm$ because this is exactly the prescription we followed to set up the origin (see Section~\ref{secSims}).
At smaller radii, we detect the motion of the very central regions, which is typically of the order of ten per cent of the maximum circular velocity.
At large radii, the velocity of the centre of mass tends asymptotically to the average velocity of the halo with respect to the background universe.
The same pattern is observed for our galaxy and cluster simulations, at the three redshifts considered.

The behaviour of the centre of mass position is at first sight more striking: rather than converging to an asymptotic value, the offset increases roughly linearly with radius, up to scales that can be as large as the box size for the largest systems.
Again, the same pattern is common to all systems, at all epochs, although the change in normalization with redshift seems to indicate that $\Rm$ is probably not the relevant scaling radius.
Obviously, the offset in position should stay constant beyond the scale of homogeneity, but this is larger than the $80\hMpc$ of the simulation box.
What Figure~\ref{figOffset} shows is that, on the scales probed by our numerical experiments, the universe is not entirely homogenous, and the tidal field of the large-scale structure is not negligible \citep[see e.g.][]{Martinez-Vaquero+07}.
One should bear in mind, however, that the finite box size and periodic boundary conditions severely limit our ability to sample the Fourier modes of the primordial fluctuations on large scales.
For $L\sim80\hMpc$, the amplitude of the CDM power spectrum is not negligible, and therefore it is possible that our results on such scales are not representative of the real universe.

Radial velocity profiles are plotted in Figure~\ref{figVr}.
Although there is, in an approximate sense, a general pattern common to all objects, at all epochs (an inner region in equilibrium, an infall region, and a region still expanding), the details depend a lot on the scale of the halo and the redshift of observation.
In units of $\Rm$, the turn-around radius of all systems stays approximately constant at $R_{\rm ta}\sim5\Rm$, but we confirm the claims by \citet{Prada+06} that, at $z=0$, the infall region is effectively absent in galaxy-mass haloes.
At earlier times ($z=1$), the progenitors of galaxies display an infall pattern similar to that of clusters at $z=0$, while the progenitors of present-day clusters feature much higher infall velocities.
In both cases, the equilibrium region where $v\rr\approx0$ is closer to $\Rm$ than to the virial radius.
The importance of infall becomes even more dramatic at $z=5$.
Maximum infall velocities are found at $r\sim2\Rm$ (about or within the virial radius of the halo), with magnitudes comparable to the maximum circular velocity of the object.
The transition to equilibrium takes place at $r\sim\Rm$ for proto-galaxies and $r\sim0.3\Rm$ for proto-clusters, relatively close to our resolution limit.

We measure the total angular momentum of the halo in terms of the spin parameter, defined as in \citet{Bullock+01_j},
\be
\lambda(r) = \frac{J}{\sqrt{2GM^3r}} = \frac{1}{\sqrt{2}} \frac{v_{\rm tan}}{\Vc}
\ee
where $v_{\rm tan}=J/Mr$ is the average tangential velocity within radius $r$.
The distribution of $\lambda$ in our 14 haloes is consistent with the average value $\lambda_0\sim0.035$ and the spread $\Delta\ln(\lambda)\sim0.6$ that are usually reported at the virial radius.
Indeed, we find (see Figure~\ref{figSpin}) that such values are roughly representative of the bulk rotation of the halo at all radii, albeit there may be large deviations in the individual profiles.
Nonetheless, we also observe a systematic trend in most objects, as well as on the average profiles: the spin parameter increases with radius until it reaches a maximum value, and then it decreases again towards the outer parts.
The intensity and location of this maximum are strongly dependent on epoch, and perhaps on the mass of the object, the maximum being stronger and closer to the centre as we consider more massive systems at higher redshift.

\subsection{Random motion}

In general, the average bulk motion of the halo is much slower than the velocity of individual dark matter particles.
The latter, which determines the orbital structure and therefore the dynamics of the system, is dominated by random motions.

Radial velocity dispersion profiles are plotted in Figure~\ref{figSigmar}.
Again, we observe a remarkable degree of universality in terms of $\Rm$ and $\Vm$, with a scatter of about 20 per cent with respect to the average behaviour within $\Rm$.
Neither the scale of the object nor the cosmic epoch seem to have a significant impact on the results, except for a slight steepening outside $\Rm$ at high redshift.
The radial velocity dispersion is well described by the R04 and DM05 models in the inner regions, but both of them seem to be systematically higher than our measurements in the outskirts of haloes (near the virial radius).
We note, however, that the discrepancy is only of the order of one sigma, and therefore more statistics are required to verify whether it is real or not.

The anisotropy parameter, defined as
\be
\beta= 1 - \frac{\sigma_{\rm tan}^2}{2\sigma\rr^2}
\ee
relates the radial and tangential components of the velocity dispersion.
For an isotropic system, $\beta=0$, while $\beta=1$ and $\beta=-\infty$ correspond to purely radial and circular orbits, respectively.
Our results, shown on the top panels of Figure~\ref{figBeta}, are well reproduced by both R04 and DM05 models within $\Rm$.
Most of the large scatter observed arises from noise in the numerical evaluation of $\beta$, but part of it (especially in the outer regions) is also due to the environment of individual haloes.
For all object types except galaxies at $z=0$ (the only ones that do not display an infall region), the anisotropy parameter attains a maximum around $\Rm$, and the decline towards larger radii seems to be sharper at high redshift than at the present time.

It has been suggested \citep{HansenMoore06,HansenStadel06} that a linear relation exists between the anisotropy parameter $\beta$ and the local value of the logarithmic slope of the density profile, $\gamma$.
As can be seen on the bottom panels of Figure~\ref{figBeta}, our data are compatible with such a relation.
A least-squares fit of the form
\be
\beta = A +B\gamma
\ee
over the range $-2.5<\gamma<-1$ yields $A=0.016\pm0.016$ and $B=-0.085\pm0.009$, with a correlation coefficient $r=-0.97$.
In this region, such a fit does not deviate significantly from the predictions of the R04 and DM05 models nor the results of \citet{HansenMoore06,HansenStadel06}, although the slope and intercept we obtain are different from the values quoted by these authors, $A=-0.16$ and $B=-0.2$.
Obviously, the extrapolation outside the fitted range (which we would strongly discourage) leads to substantially different results in both directions.

\begin{figure*}
\includegraphics[width=5.5cm]{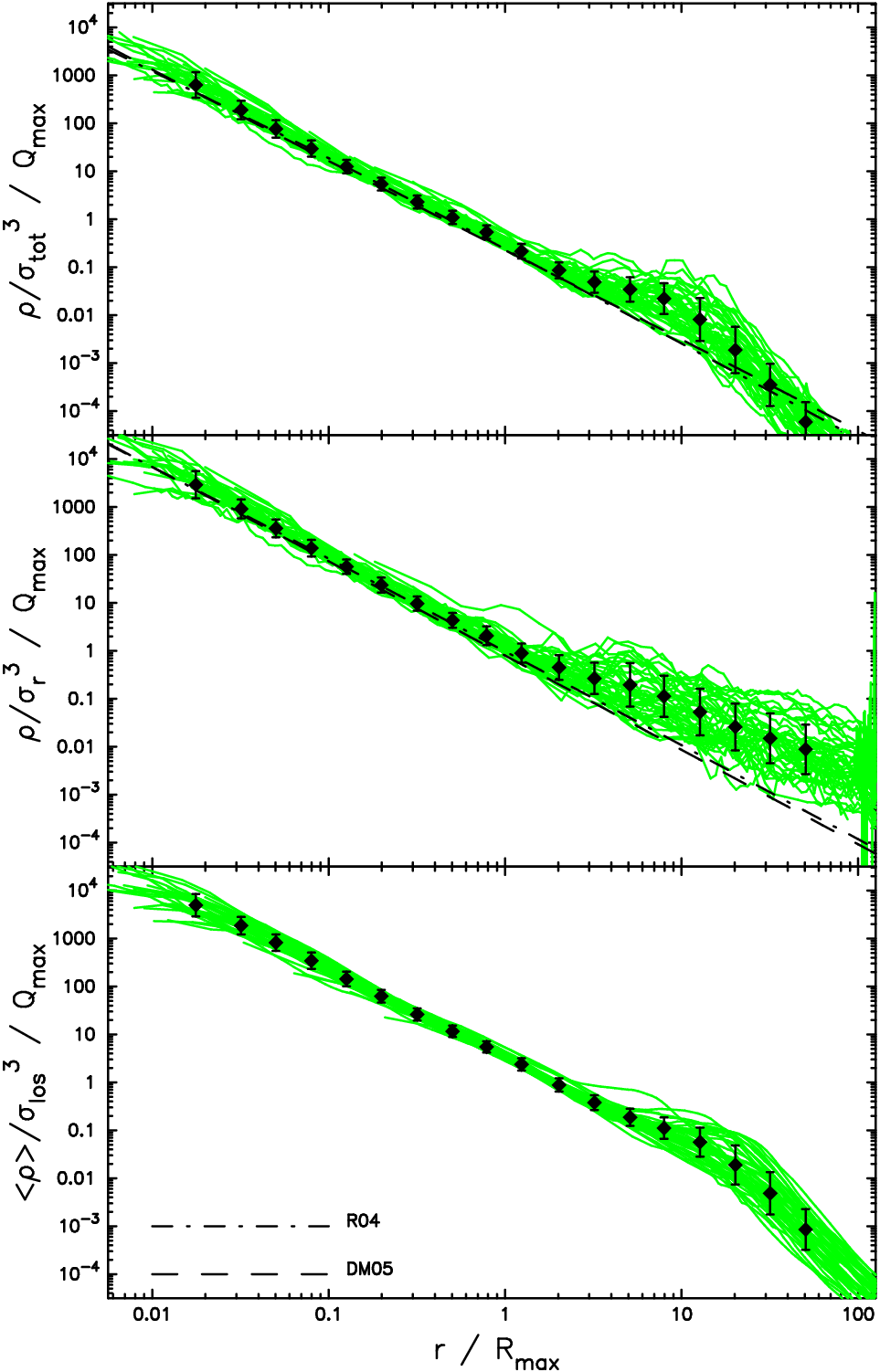}
\hfill
\includegraphics[width=5.5cm]{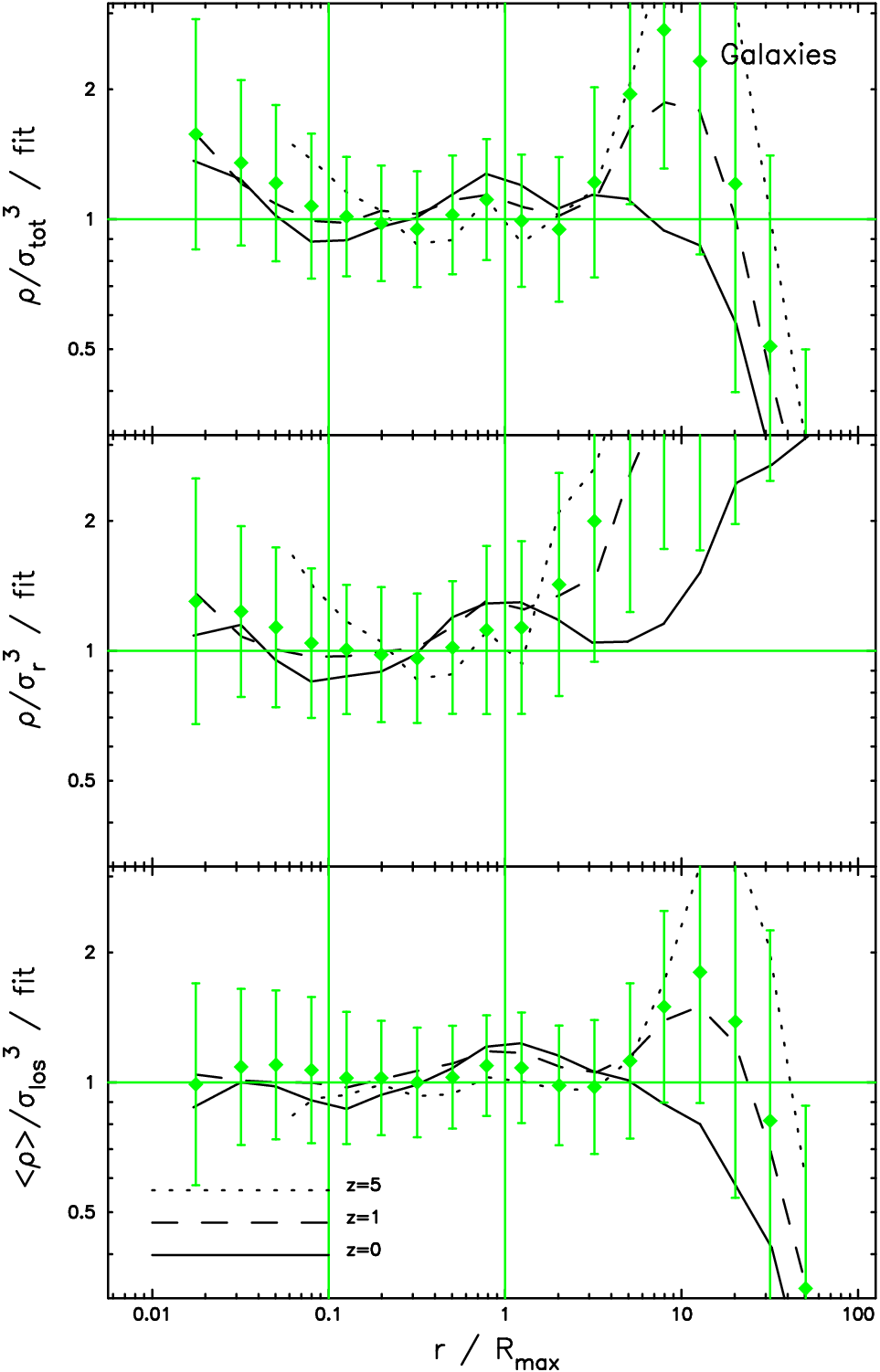}
\hfill
\includegraphics[width=5.5cm]{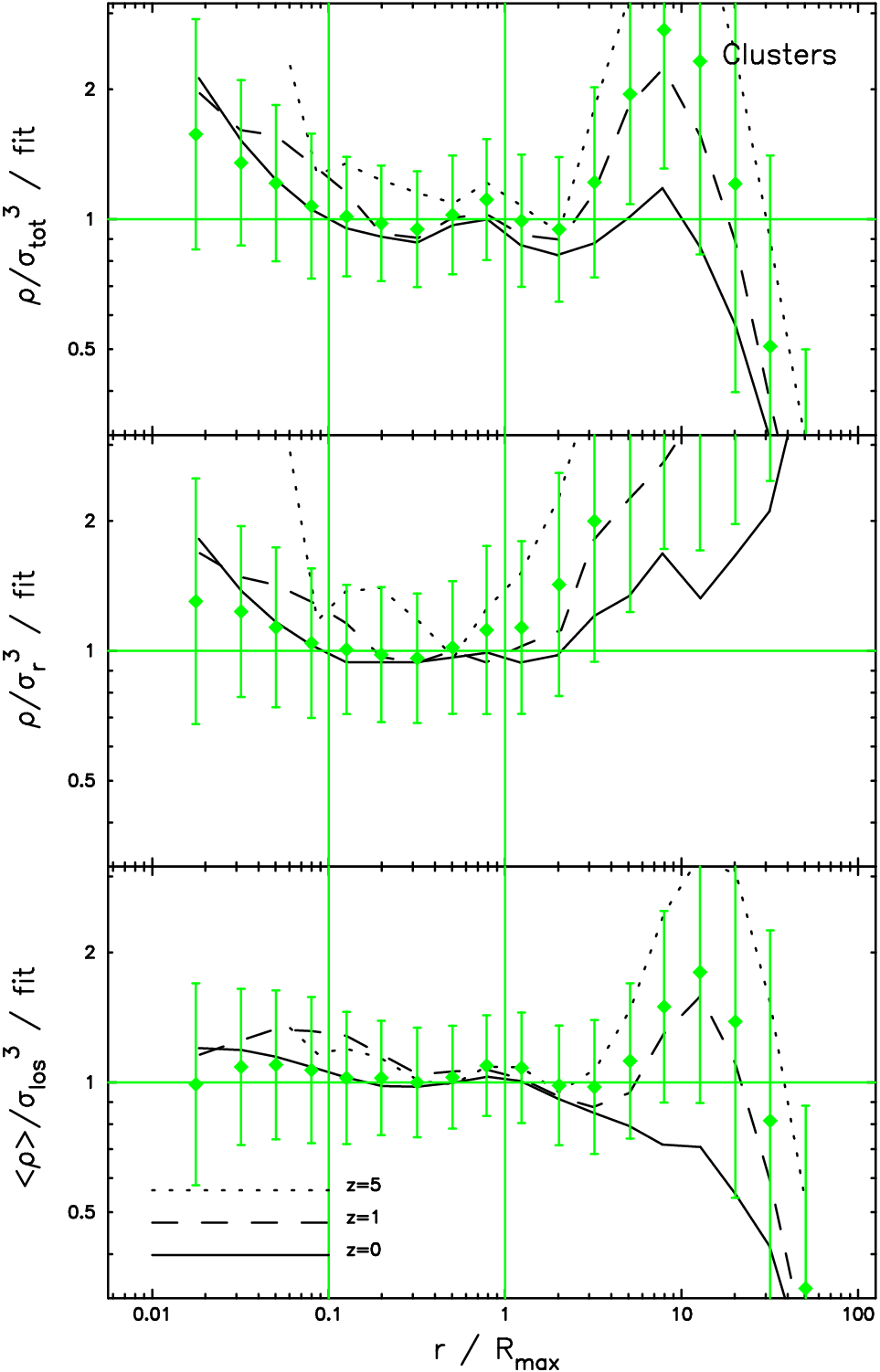}
\caption
{
  Coarse-grained phase-space density, according to three slightly different definitions (see text for details).
  Line styles are the same as in Figure~\ref{figVc}.
  \Revised{
    Middle and right panels show the residuals with respect to the best-fitting power law, considering the whole sample.
    Vertical lines indicate the radial region used for the fit.
  }
}
\label{figQ}
\end{figure*}

Finally, we investigate the coarse-grained phase-space density profile of our haloes in Figure~\ref{figQ}, comparing three different definitions of this quantity.
On the top panels, we plot
\be
q_{\rm tot}(r)=\frac{\rho}{\sigma_{\rm tot}^3}
\ee
where $\sigma_{\rm tot}^2=\sigma\rr^2+\sigma_{\rm tan}^2=(3-2\beta)\sigma\rr^2$ denotes the total velocity dispersion at radius $r$.
On the middle panels, the radial velocity dispersion is used instead,
\be
q\rr(r)=\frac{\rho}{\sigma\rr^3} = \left(3-2\beta\right)^{3/2} q_{\rm tot}
\ee
and on the bottom panel, the coarse-grained phase-space density is defined in terms of cumulative quantities,
\be
q_{\rm los}(r)=\frac{\left<\rho\right>}{\sigma_{\rm los}^3}
\ee
where $\left<\rho\right>=3M/4\pi r^3$ is the average density within radius $r$ and
$\sigma_{\rm los}^2 = \frac{1}{3M} \int ( v^2 - \left<v\right>^2 ) 4\pi r^2 \rho\ \dd r$ is the average line-of-sight velocity dispersion.

Our results indicate that the coarse-grained phase-space density profile, defined in any of the aforementioned ways, can be approximately described by a power law, in agreement with previous work.
Considering data in the range $0.1<r/\Rm<1$, we find
\be
\frac{q_i(r)}{Q\mm}= K_i \, r^{-\alpha_i}
\ee
with best-fitting exponents $\alpha_{\rm tot}= 1.77 \pm 0.04$, $\alpha_{\rm r}= 1.87 \pm 0.03$, and $\alpha_{\rm los}= 1.82 \pm 0.01$ for the quantities $q_{\rm tot}$, $q\rr$ and $q_{\rm los}$, respectively.
The corresponding normalizations are $K_{\rm tot}=10^{-0.50\pm0.03}$, $K_{\rm r}=10^{0.07\pm0.02}$, and $K_{\rm los}=10^{0.51\pm0.01}$, and the quantity $Q\mm$ is defined as
\be
Q\mm= \frac{\rho\mm}{\Vm^3} = \frac{3}{4\pi G\ \Vm\Rm^2}
\ee
in all cases.
\Revised{
A qualitatively similar behaviour is obtained for $\rho/\sigma_{\rm tan}^3$, with $\alpha_{\rm tan}=1.70\pm0.05$ and $K_{\rm tan}=10^{-0.15\pm0.03}$.
}

Regardless of the exact definition of $q$, both sets of simulations (galaxy- and cluster-size objects) follow the same power law, and cosmic evolution can only be appreciated at very large radii (usually well beyond the virial radius)\footnote{\Revised{
The apparent deviations at $r<0.01\Rm$ should be taken with a grain of salt, since they might well be related to our choice of the origin of velocities.
}}.
Therefore, these power laws can be used to provide a simple, phenomenological description of the structure of dark matter haloes.
However, as in the case of the anisotropy profile (or any of the other relations investigated here), we would strongly advise against assuming that any given analytic fit is an \emph{exact} description, or that it can be safely extrapolated outside its range of validity.

For instance, the first two definitions of the coarse-grained phase space density could only follow exactly a power law if the anisotropy was constant at all radii, which is definitely not the case, or it varied according to
\be
\beta(r)=\frac{3-(r/R_1)^{-\epsilon}}{2}
\label{eqBetaQ}
\ee
where the exponent $\epsilon=\alpha_{\rm r}-\alpha_{\rm tot}\simeq0.1$ would depend on the precise values of the logarithmic slopes of the $q\rr$ and $q_{\rm tot}$ profiles, and $R_1$ would be a characteristic radius at which $\beta=1$ and $q\rr=q_{\rm tot}$.
Expression~(\ref{eqBetaQ}) does indeed provide a good fit to our simulated data, but one should not conclude from there that the innermost orbits are perfectly circular, i.e. $\beta(0)=-\infty$, because there is no \emph{a priori} reason to think that $q\rr(r)$ and $q_{\rm tot}(r)$ ought to be exact power laws.

  \section{Conclusions}
  \label{secConclus}

In this paper, we have studied the internal structure of dark matter haloes using high-resolution cosmological $N$-body simulations of individual objects.
Our sample consists of six dark matter haloes on galactic scales and eight systems that correspond to galaxy groups and clusters at the present day.
These objects have been randomly selected, without imposing any condition on their dynamical state.
We have considered the radially-averaged profiles of cumulative mass and circular velocity, dark matter density and its logarithmic slope, position and velocity offsets of the centre of mass on different scales, average radial velocity and spin parameter, radial velocity dispersion and anisotropy parameter, and coarse-grained phase-space density.

Our results show that the all these profiles display a similar structure when expressed in terms of the maximum circular velocity of the halo, $\Vm$, and the radius at which it is attained, $\Rm$.
Within $\Rm$, this structure is independent on object scale and cosmic epoch, and it is well fit by several analytical models that have previously been proposed to describe the numerical data at $z=0$ in terms of other scaling parameters.
\Revised{
This indicates that the equilibrium properties of dark matter haloes are independent of the individual merger histories, and therefore they must be set by some physical process or processes (such as violent relaxation, phase mixing, radial orbit instability, etc.) whose outcome is similar for all objects.
In particular, the present study rules out theoretical models that predict mass or redshift dependence of the dynamical structure of dark matter haloes.
}

The radius $\Rm$ gives a more conservative indication of the region where a given object can be considered in equilibrium than other measurements, such as the virial radius.
This is particularly important at high redshift, where significant radial infall (up to fifty per cent of $\Vm$) may take place within the latter.
Outside $\Rm$, not only the radial velocity profile, but also the mass distribution, the spin, or the anisotropy parameter display strong systematic dependences on mass and/or redshift.
\Revised{
It is in this outer region where the system retains some memory of the primordial initial conditions and displays more obvious signatures of its current environment.
}

Finally, we provide fits for two relations between position- and velocity-related quantities.
We corroborate the roughly linear relation between the anisotropy parameter and the local logarithmic slope of the density profile, although the best-fitting zero-point and proportionality constant we obtain differ from the values originally proposed.
We also compute the slope and normalization of the coarse-grained phase-space density profile, which is found to be \Revised{fairly} well described by a pure power law, according to three different definitions.
\Revised{
These two quantities, as well as all the others considered in the present study, involve integrals of the underlying distribution function (the fine-grained, six-dimensional phase-space density) over the velocity.
The distribution function itself is very difficult to measure accurately \citep{Arad+04,AscasibarBinney05,SharmaSteinmetz06}, but it should certainly be the aim of future work in order to fully characterize the structure of dark matter haloes.
Promising steps in that direction have been taken by e.g. \citet{Hansen+06}, who consider the velocity distribution function within different bins in potential energy, and \citet{Wojtak+/08}, who propose a simple parametric form in terms of the energy, angular momentum, and anisotropy of particle orbits.
}

We would like to stress, though, that caution must be exercised when interpreting the results of phenomenological studies (like the present one).
In particular, analytical fits are an extremely useful tool that provides an \emph{approximate} description of dark matter haloes, but they constitute by no means a physical law.
Extrapolating outside the appropriate validity range of or using a fit as an exact description can lead to misleading conclusions, not only quantitatively, but also on a qualitative level.
More generally, we do not share the widespread opinion that the formation and evolution of dark matter haloes is nowadays well understood.
Even if we could provide a good phenomenological prescription of their dynamical structure in terms of empirical relations, there is an absolute lack of understanding of their physical origin.
Open questions are, for example, how many parameters are actually needed to describe a dark matter halo?
Which relations (if any) are indeed fundamental?
What physical processes drive the evolution of dark matter haloes, and why do they lead to a `universal' equilibrium state?
We hope that the results presented here help in providing a benchmark to test any physically-motivated model that is able to address these questions.


\section*{Acknowledgments}

We would like to thank \Revised{the referee for his/her insightful comments, as well as} Y. Hoffman and A. Klypin for useful discussions related to the present work.
The numerical simulations on which it is based were performed at the LRZ Munich, NIC J\"ulich, and NASA Ames.


 \bibliographystyle{mn2e}
 \bibliography{references}


\end{document}